\documentclass[twocolumn]{aastex631}

\received{April 2022}
\revised{June 2022}
\accepted{July 2022}
\submitjournal{AJ}
\shorttitle{}
\shortauthors{}
\shortauthors{Swimmer et al.}
\graphicspath{{./}{figures/}}
\begin{document}

\title{SCExAO and Keck Direct Imaging Discovery of a Low-Mass Companion Around the Accelerating F5 Star HIP 5319\footnote{Based in part on data collected at Subaru Telescope, which is operated by the National Astronomical Observatory of Japan. Some of the data presented herein were obtained at the W. M. Keck Observatory, which is operated as a scientific partnership among the California Institute of Technology, the University of California and the National Aeronautics and Space Administration. The Observatory was made possible by the generous financial support of the W. M. Keck Foundation. This work makes use of observations from the Las Cumbres Observatory global telescope network (LCOGT).}}

 \correspondingauthor{Noah Swimmer}
 \email{nswimmer@ucsb.edu}
 \author[0000-0001-5721-8973]{Noah Swimmer}
 \affiliation{Department of Physics, University of California, Santa Barbara, Santa Barbara, California, USA}
 \author[0000-0002-7405-3119]{Thayne Currie}
 \affiliation{Subaru Telescope, National Astronomical Observatory of Japan, 
650 North A`oh$\bar{o}$k$\bar{u}$ Place, Hilo, HI  96720, USA}
\affiliation{Department of Physics and Astronomy, University of Texas-San Antonio, San Antonio, TX, USA}
 \affiliation{NASA-Ames Research Center, Moffett Blvd., Moffett Field, CA, USA}
 \affiliation{Eureka Scientific, 2452 Delmer Street Suite 100, Oakland, CA, USA}
 \author[0000-0002-4787-3285]{Sarah Steiger}
 \affiliation{Department of Physics, University of California, Santa Barbara, Santa Barbara, California, USA}
 \author{G. Mirek Brandt}
 \affiliation{Department of Physics, University of California, Santa Barbara, Santa Barbara, California, USA}
 \author{Timothy D. Brandt}
 \affiliation{Department of Physics, University of California, Santa Barbara, Santa Barbara, California, USA}
 \author{Olivier Guyon}
 \affiliation{Subaru Telescope, National Astronomical Observatory of Japan, 
 650 North A`oh$\bar{o}$k$\bar{u}$ Place, Hilo, HI  96720, USA}
 \affil{Steward Observatory, The University of Arizona, Tucson, AZ 85721, USA}
 \affil{College of Optical Sciences, University of Arizona, Tucson, AZ 85721, USA}
 \affil{Astrobiology Center, 2-21-1, Osawa, Mitaka, Tokyo, 181-8588, Japan}
 \author{Masayuki Kuzuhara}
 \affiliation{Astrobiology Center, 2-21-1, Osawa, Mitaka, Tokyo, 181-8588, Japan}
 \affiliation{National Astronomical Observatory of Japan, 2-21-2, Osawa, Mitaka, Tokyo 181-8588, Japan}
 \author{Jeffrey Chilcote}
 \affiliation{Department of Physics, University of Notre Dame, South Bend, IN, USA}
 \author{Taylor Tobin}
 \affiliation{Department of Physics, University of Notre Dame, South Bend, IN, USA}
 \author{Tyler D. Groff}
 \affiliation{NASA-Goddard Space Flight Center, Greenbelt, MD, USA}
 \author{Julien Lozi}
 \affiliation{Subaru Telescope, National Astronomical Observatory of Japan, 
 650 North A`oh$\bar{o}$k$\bar{u}$ Place, Hilo, HI  96720, USA}
 \author[0000-0002-4272-263X]{John I. Bailey, III}
 \affiliation{Department of Physics, University of California, Santa Barbara, Santa Barbara, California, USA}
 \author{Alexander B. Walter}
 \affiliation{Jet Propulsion Laboratory, California Institute of Technology, Pasadena, California 91125, USA}
 \author{Neelay Fruitwala}
 \affiliation{Department of Physics, University of California, Santa Barbara, Santa Barbara, California, USA}
 \author[0000-0003-3146-7263]{Nicholas Zobrist}
 \affiliation{Department of Physics, University of California, Santa Barbara, Santa Barbara, California, USA}
 \author[0000-0002-0849-5867]{Jennifer Pearl Smith}
 \affiliation{Department of Physics, University of California, Santa Barbara, Santa Barbara, California, USA}
 \author{Gregoire Coiffard}
 \affiliation{Department of Physics, University of California, Santa Barbara, Santa Barbara, California, USA}
 \author{Rupert Dodkins}
 \affiliation{Department of Physics, University of California, Santa Barbara, Santa Barbara, California, USA}
 \author[0000-0001-5587-845X]{Kristina K. Davis}
 \affiliation{Department of Physics, University of California, Santa Barbara, Santa Barbara, California, USA}
 \author{Miguel Daal}
 \affiliation{Department of Physics, University of California, Santa Barbara, Santa Barbara, California, USA}
 \author{Bruce Bumble}
 \affiliation{Jet Propulsion Laboratory, California Institute of Technology, Pasadena, California 91125, USA}
 \author{Sebastien Vievard}
 \affiliation{Subaru Telescope, National Astronomical Observatory of Japan, 
650 North A`oh$\bar{o}$k$\bar{u}$ Place, Hilo, HI  96720, USA}
 \author[0000-0002-9372-5056]{Nour Skaf}
 \affiliation{Subaru Telescope, National Astronomical Observatory of Japan, 
650 North A`oh$\bar{o}$k$\bar{u}$ Place, Hilo, HI  96720, USA}
\affiliation{LESIA, Observatoire de Paris, Univ.~PSL, CNRS, Sorbonne Univ., Univ.~de Paris, 5 pl. Jules Janssen, 92195 Meudon, France}
\affiliation{Department of Physics and Astronomy, University College London, London, United Kingdom}
 \author{Vincent Deo}
 \affiliation{Subaru Telescope, National Astronomical Observatory of Japan, 
 650 North A`oh$\bar{o}$k$\bar{u}$ Place, Hilo, HI  96720, USA}
 \author[0000-0001-5213-6207]{Nemanja Jovanovic}
 \affiliation{Department of Astronomy, California Institute of Technology, 1200 E. California Blvd.,Pasadena, CA, 91125, USA}
 \author{Frantz Martinache}
 \affiliation{Universit\'{e} C\^{o}te d'Azur, Observatoire de la C\^{o}te d'Azur, CNRS, Laboratoire Lagrange, France}
 \author{Motohide Tamura}
 \affil{Astrobiology Center, 2-21-1, Osawa, Mitaka, Tokyo, 181-8588, Japan}
 \affiliation{National Astronomical Observatory of Japan, 2-21-2, Osawa, Mitaka, Tokyo 181-8588, Japan}
 \affiliation{Department of Astronomy, Graduate School of Science, The University of Tokyo, 7-3-1, Hongo, Bunkyo-ku, Tokyo, 113-0033, Japan}
 \author{N. Jeremy Kasdin}
 \affiliation{University of San Francisco, San Francisco, CA 94118}
 \author[0000-0003-0526-1114]{Benjamin A. Mazin}
 \affiliation{Department of Physics, University of California, Santa Barbara, Santa Barbara, California, USA}

%% Note that the \and command from previous versions of AASTeX is now
%% depreciated in this version as it is no longer necessary. AASTeX 
%% automatically takes care of all commas and "and"s between authors names.

%% AASTeX 6.31 has the new \collaboration and \nocollaboration commands to
%% provide the collaboration status of a group of authors. These commands 
%% can be used either before or after the list of corresponding authors. The
%% argument for \collaboration is the collaboration identifier. Authors are
%% encouraged to surround collaboration identifiers with ()s. The 
%% \nocollaboration command takes no argument and exists to indicate that
%% the nearby authors are not part of surrounding collaborations.

%% Mark off the abstract in the ``abstract'' environment. 
\begin{abstract}
We present the direct imaging discovery of a low-mass companion to the nearby accelerating F star, HIP 5319, using  SCExAO coupled with the CHARIS, VAMPIRES, and MEC instruments in addition to Keck/NIRC2 imaging. CHARIS \textit{JHK} (1.1-2.4 $\mu$m) spectroscopic data combined with VAMPIRES 750 nm, MEC \textit{Y}, and NIRC2 $L_{\rm p}$ photometry is best matched by an M3--M7 object with an effective temperature of T=3200 K and surface gravity log(\textit{g})=5.5. Using the relative astrometry for HIP 5319 B from CHARIS and NIRC2 and absolute astrometry for the primary from \textit{Gaia} and \textit{Hipparcos} and adopting a log-normal prior assumption for the companion mass, we measure a dynamical mass for HIP 5319 B of $31^{+35}_{-11}M_{\rm J}$, a semimajor axis of $18.6^{+10}_{-4.1}$ au, an inclination of  $69.4^{+5.6}_{-15}$ degrees, and an eccentricity of $0.42^{+0.39}_{-0.29}$.   However, using an alternate prior for our dynamical model yields a much higher mass of 128$^{+127}_{-88}M_{\rm J}$.   Using data taken with the LCOGT NRES instrument we also show that the primary HIP 5319 A is a single star in contrast to previous characterizations of the system as a spectroscopic binary. This work underscores the importance of assumed priors in dynamical models for companions detected with imaging and astrometry and the need to have an updated inventory of system measurements.
\end{abstract}


%% From the front matter, we move on to the body of the paper.
%% Sections are demarcated by \section and \subsection, respectively.
%% Observe the use of the LaTeX \label
%% command after the \subsection to give a symbolic KEY to the
%% subsection for cross-referencing in a \ref command.
%% You can use LaTeX's \ref and \label commands to keep track of
%% cross-references to sections, equations, tables, and figures.
%% That way, if you change the order of any elements, LaTeX will
%% automatically renumber them.
%%
%% We recommend that authors also use the natbib \citep
%% and \citet commands to identify citations.  The citations are
%% tied to the reference list via symbolic KEYs. The KEY corresponds
%% to the KEY in the \bibitem in the reference list below. 

\section{Introduction} \label{sec:intro}
%Over the past two decades direct imaging detections of exoplanets around young, nearby stars have increased with facility adaptive optics (AO) systems and now extreme adaptive optics systems \citep{Marois2008, Marois2010, Kuzuhara2013, Currie2015, Macintosh2015,Chauvin2017,Keppler2018,Currie2022a}.
%The majority of these detections have been of companions at separations greater than $\rho\sim0\farcs{}4$ and correspond to separations of 10-150 au from their host stars \citep[][]{Nielsen2019,Vigan2021,Currie2022b}. 
Over the past two decades, both facility adaptive optics (AO) systems and now extreme AO systems have provided numerous images of planets and low-mass brown dwarfs around nearby stars \citep[e.g.][]{Marois2008, Marois2010, Thalmann2009, Carson2013, Kuzuhara2013, Currie2014, Macintosh2015,Konopacky2016,Chauvin2017,Cheetham2018,Keppler2018,Currie2022a}.   The majority of discoveries draw from \textit{blind} (or ``unbiased") surveys, where targets are selected based on age and distance \citep[e.g.][]{Desidera2021}.  However, these same surveys show that occurrence rates of detectable moderate-to-wide separation planets and brown dwarf companions is low, $\sim$a few percent around FGK stars \citep{Vigan2021,Nielsen2019,Currie2022b}.

%More recently, second generation extreme AO systems such as the Subaru Coronagraphic Extreme Adaptive Optics system \citep[SCExAO;][]{Jovanovich2015b, Ahn2021} and the Magellan Extreme Adaptive Optics system \citep[MagAO-X;][]{Males2020} have enabled deeper contrasts at separations less than $\rho\lesssim0\farcs{}4$. These deeper contrasts - in conjunction with noise-reducing techniques such as Stochastic Speckle Discrimination \citep[SSD;][]{Walter2019} and Angular, Spectral, and Reference Differential Imaging \citep[ADI, SDI, RDI;][]{Marois2006, Marois2000, Ruane2019} - have led to the discovery of companions with orbital radii less than 10 au \citep[][]{Chilcote2021, Steiger2021}. By continuing to drive to deeper contrasts at narrower separations we will be able to directly image companions with orbits closer to their host stars and around host stars at greater distances from earth.

Recent work has demonstrated the success instead of dynamics-selected direct imaging surveys, specifically using precision astrometry from the \textit{Gaia} and \textit{Hipparcos} satellites in the \textit{Hipparcos-Gaia Catalog of Accelerations} to identify stars showing a proper motion anomaly -- i.e. an astrometric acceleration -- likely due to an unseen low-mass companion \citep{vanLeeuwen2007,GaiaDR2,Gaia2021,BrandtHGCA2021}.   Direct imaging of targets showing an acceleration from HGCA have revealed white dwarfs \citep{Bonavita2020}, low-mass stars \citep{Steiger2021,Chilcote2021}, moderate-to-low mass brown dwarfs \citep{Currie2020,Bowler2021,Bonavita2022,Kuzuhara2022}, and soon will show planets.
%now planets \citep{Currie2022c}. 

Jointly analyzing absolute astrometry of the star from HGCA and relative astrometry of the imaged companion with Markov-Chain Monte Carlo (MCMC) codes like \texttt{orvara} \citep{orvara2021} can provide strong constraints on the companion's dynamical mass and orbit \citep[e.g.][]{Brandt2021}.   To derive these constraints, MCMC codes require input priors for the orbital parameters, primary mass, and companion mass(es).   Typical orbital priors include a log-normal distribution in semimajor axis ($p(a) \propto 1/a$), uniform prior in inclination ($p(i) \propto sin(i)$), gaussian prior in primary mass, and log-normal prior in companion mass ($p(M_{2}) \propto 1/M_{2}$) \citep[e.g.][]{Kuzuhara2022}.

While the above orbital priors are long regarded as standard in MCMC modeling \citep[e.g.][]{Blunt2020}, the most appropriate companion prior may differ.  The initial mass function for companions near the substellar to stellar boundary exhibits a more gaussian-like distribution \citep[e.g.][]{Chabrier2003}: i.e. a turnover in the mass function near the hydrogen-burning limit.   Ancillary system properties -- e.g. age, primary and companion spectral type, etc. -- also are often used to inform adopted priors but may derive from heterogeneously-sourced data.

Here, we report the direct imaging discovery of HIP 5319 B: a low mass -- potentially substellar -- companion around the F-type star HIP 5319 A using the SCExAO Subaru Coronagraphic Extreme Adaptive Optics system \citep[SCExAO;][]{Jovanovich2015b, Ahn2021} coupled with the MKID Exoplanet Camera \citep[MEC;][]{Walter2020}, the Visible Aperture Masking Polarimetric Imager for Resolved Exoplanetary Structures \citep[VAMPIRES;][]{Norris2015}, the Coronagraphic High Angular Resolution Imaging Spectrograph \citep[CHARIS;][]{Groff2016}, and the NIRC2 camera on the Keck II telescope.  
%In addition to being another low-mass companion discovered through a joint direct imaging and astrometry survey, 
HIP 5319 B illustrates the sensitivity of adopted priors for companion mass for parameters derived from jointly modeling direct imaging and astrometric data and the need to verify ancillary information about the system -- e.g. binarity, age, rotation -- in direct imaging + astrometric surveys.

% The companion is responsible for the astrometric acceleration of its primary star as identified in the HGCA.   Modeling astrometric measurements suggest a best-fit dynamical mass of $30^{+35}_{-11}M_{\rm Jup}$ or 128$^{+127}_{-88}M_{\rm J}$, depending on the assumed priors. The spectroscopy from CHARIS combined with photometry from MEC, NIRC2, and VAMPIRES are most consistent with a spectral type M3-M7 from spectral analysis.

\section{Stellar Properties and Observations}
\subsection{HIP 5319 A Basic Properties}

HIP 5319 ($^{\star}$78 Psc) is an F5IV spectral class star \citep{BoroSaikia2018} at $d=$42.93$\pm$0.06 pc (\citet{GaiaMission}, \citet{GaiaDR2}).  Banyan-$\Sigma$ \citep{Gagne2018} shows no evidence that the system is a member of any moving group or young association.   It has previously been identified as an RS CVn binary star by \citet{Fleming1989}, who measured a projected rotation rate of \textit{v}sin(\textit{i})=68$\pm$20.5 km/s and x-ray luminosity of $L_{x}=9.2\pm3.7\times10^{28}$ erg/s.  

\subsubsection{System Age}
Evidence informing the HIP 5319 system's age is complex.   On one hand, HIP 5319 has an extreme level of chromospheric activity (log($R^\prime_{HK}$)= $-$4.016) as measured by Calcium II H and K lines, which tracks the strength of the emission at the cores of the 2 lines \citep{BoroSaikia2018}.  The chromospheric index easily exceeds values for stars in the Pleiades and Hyades associations and is comparable or higher to the stars in the Scorpius-Centaurus (Sco-Cen) association \citep[][]{MamajekHillenbrand2008, Pecaut2013}. Its Hertzsprung-Russell diagram position in \textit{Gaia} color-magnitude space ($M_{G}$ vs $G_{BP}-G_{RP}$=2.97, 0.54) lies between the Pleiades and Hyades, which is consistent with either a main sequence star between 115 and $\sim$800 Myr, respectively \citep[][]{Gossage2018}, or a pre-main sequence star much younger than the Pleiades.   Based on its activity, \citet{StanfordMoore2020} estimate a young age of 75$^{+492}_{-63.5}$ $Myr$.

HIP 5319 was also observed by the Transiting Exoplanet Survey Satellite \citep[TESS;][]{TESS_Mission} and has 2-minute cadence photometry for one sector. This observation may be too short to show spots reliably, but it does show pulsations with a period of just less than 1 day\footnote{Accessed via \url{https://mast.stsci.edu/portal/Mashup/Clients/Mast/Portal.html}}. It was also observed once by the International Ultraviolet Explorer (IUE) during IUE Program ID: CB401 \citep[Stellar Chromospheres;][]{Blanco1982}. In the spectrum from IUE\footnote{Accessed via \url{https://archive.stsci.edu/iue/obtaining.html}}, HIP 5319 A shows strong emission from the Lyman $\alpha$ line. These two data points show signs that the primary might be chromospherically active, though follow up observation is required to determine the nature of this activity.

On the other hand, RS CVn binaries -- of which HIP 5319 is claimed to be an example -- typically have orbital periods less than 14 days and show high levels of chromospheric activity via strong emission in Calcium II H and K lines, and have a hotter component of spectral type F or G \citep[][]{Montesinos1988}.  Multiple sources have reported \textit{v}sin(\textit{i}) values with significant scatter, which may suggest binarity: 125 km/s \citep[][]{Danziger1972}, 68$\pm$20.5 km/s \citep[][]{Fleming1989}, 36.4$\pm$4.8 km/s \citep[][]{deMedeiros1999}, 35 km/s \citep[][]{Nordstrom2004}, and 41.5 km/s (\citet{Glebocki2005}, \citet{Glebocki2005Catalog}).   The fractional x-ray luminosity of the star is log($L_{x}$/$L_{bol}$)$\sim$-4.9 \citep[][]{Gioia1990, Favata1995}, almost two orders of magnitude less than a typical pre-main sequence star, which would have values of log($L_{x}$/$L_{bol}$)$\sim$-3.2 for fractional x-ray luminosity \citep[][]{Preibisch2005}, respectively. Other authors have estimate the age of the star using isochrones and have found values of $1.6^{+0.3}_{-0.4}$ Gyr \citep[][]{Holmberg2009} and 1.07-1.23 Gyr using Padova and BASTI models \citep[][]{Casagrande2011}.
%lending credence to its identification as an older, single star.

Ultimately, the conflicting identifications of the HIP 5319 primary as either a young, chromospherically active star or an older star whose Ca II HK emission is due to a close binary will have significant implication on the understanding of the stellar system and interpretation of any of its companions' properties. If there is not significant HK emission and little evidence of binarity then the higher age estimate is likely the correct one, which will anchor the interpretation of its companion. Therefore, in addition to performing a direct imaging search for such a binary companion, a spectroscopic study of the primary with a high resolution spectrograph is necessary to disentangle the possible identities of the star and settle on the correct interpretation. This will be discussed further in sections \ref{sub:primary_single_star} and \ref{sub:ca_hk}.

% The \textit{Hipparcos-Gaia} Catalog of Accelerations reports a $\chi^{2}=$171.04: evidence of a 12.9-$\sigma$ significant acceleration of the primary with 2 degrees of freedom \citep[][]{BrandtHGCA2021}). \textbf{The statistically significant acceleration of HIP 5319 is suggestive of the presence of a previously unseen low-mass companion. Therefore we chose to observe this target in an attempt to uncover any previously unimaged low-mass companions around this accelerating star, following a similar method of target selection as in \citet{Currie2020} and \citet{Steiger2021}.}

\subsubsection{Evidence for An Astrometric Acceleration}
The \textit{Hipparcos-Gaia} Catalog of Accelerations reports a $\chi^{2}=$171.04: evidence of a 12.9-$\sigma$ significant acceleration of the primary with 2 degrees of freedom \citep[][]{BrandtHGCA2021}. The statistically significant acceleration of HIP 5319 is suggestive of the presence of a previously unseen low-mass companion at a $\gtrsim$10 au scale.   HIP 5319 was not known to have a wide-separation binary companion that could plausibly be source of this acceleration.

Therefore we chose to observe this target in an attempt to uncover any previously unimaged low-mass companions around this accelerating star, following a similar method of target selection as in \citet{Currie2020} and \citet{Steiger2021}.

% NS- below is likely no longer needed since B-V vs log(R'HK) probably doesn't get us a great age estimation for such an early type F-star
% \begin{figure}
%     \centering
%     \includegraphics[scale=0.5]{young_star_cluster_BVvRHK_withHip5319.JPG}
%     \caption{\textcolor{red}{Placeholder, from Thayne. HIP 5319 shown as a blue star.}}
%     \label{fig:youngClusterWHipAdded}
% \end{figure}

%TC- below is probably not needed
%\begin{figure}
%%    \centering
 %   \includegraphics[scale=0.5]{young_star_cluster_BVvRHK_MamajekHillebrand_withHip5319.JPG}
 %   \caption{Reproduced from \cite{MamajekHillenbrand2008} with HIP 5319 shown as a blue star. Includes Sco-Cen (not included in figure \ref{fig:youngClusterWHipAdded})}
 %   \label{fig:youngClusterFromMH}
%\end{figure}

\begin{deluxetable*}{lllllllllll}[ht!]
    \tablewidth{0pt}
    \tabletypesize{\scriptsize}
    \tablecaption{HIP 5319 Observing Log}
    \tablehead{\colhead{UT Date} & \colhead{Instrument} &  \colhead{coronagraph} & \colhead{Seeing (\arcsec{})} &{Passband} & \colhead{$\lambda$ ($\mu$m)$^{a}$} 
    & \colhead{$t_{\rm exp}$ (s)} & \colhead{$N_{\rm exp}$} & \colhead{$\Delta$PA ($^{o}$)} & \colhead{Post-Processing Strategy} }
    %\\
    %{} & {} & {} & {} & {} & {} & {} & {} & {} &    }
    \startdata
    20200731 & SCExAO/CHARIS & Lyot & 0.4--0.6  & $JHK$ & 1.16--2.37 & 30.98 & 14 & 5.3 & RDI-KLIP  \\
    --  & SCExAO/MEC & Lyot & -- & $Y$ & 0.95--1.12 & 5.0-10.0 & 61$^{b}$ & 4.6 & none \\
     20210911 & SCExAO/CHARIS & Lyot  &0.5--0.6  & $JHK$  & 1.16--2.37 & 30.98  & 8 (32)$^{c}$ & 9.9 & none \\
     -- & SCExAO/VAMPIRES & --  & -- & 750nm & 0.75 & 12.8 & 48  & 11.2 & ADI-ALOCI \\
     20220115 & Keck/NIRC2 & none & 0.6 & L$_{p}$ & 3.78 & 30 & 30 & 9.1 & RDI-KLIP \\
     20220119  & SCExAO/MEC & Lyot & 0.7 & $YJ$ & 0.95--1.4 & 15 & 49 & 3.8 & none \\
    \enddata
    \tablecomments{a) For CHARIS and MEC data, this column refers to the wavelength range.  For broadband imaging data, it refers to the central wavelength. b) Total integration time is 430 s. c) In total, we obtained 32 exposures but only 8 were retained due to substantial PSF core splitting from low-wind effect.
    }
    \label{obslog_hip5319}
    \vspace{-0.5cm}
\end{deluxetable*}

\subsection{Observations and Data Reduction}
\label{sub:observations_and_datareduction}

 \begin{deluxetable}{llllll}
    \tablewidth{\columnwidth}
    \tabletypesize{\small}
    \tablecaption{HIP 5319 LCOGT Observing Log$^{a}$}
    \tablehead{\colhead{BJD} &  \colhead{$t_{\rm exp}$ (s)} &  \colhead{SNR$^{b}$} & \colhead{RV (km/s)} & \colhead{\textit{v}sin(\textit{i}) (km/s)}}
    %\\
    %{} & {} & {} & {} & {} & {} & {} & {} & {} &    }
    \startdata
    2459600.268 & 1000 & 230 & 17.30$\pm$1.80 & 95.24$\pm$1.65 \\
    2459601.266 & -- & 237 & 17.94$\pm$1.59 & 95.63$\pm$1.59 \\
    2459605.194 & 1500 & 316 & 18.12$\pm$1.81 & 93.37$\pm$1.64 \\
    2459607.221 & 1000 & 227 & 16.07$\pm$1.23 & 95.89$\pm$1.64 \\
    2459608.227 & -- & 246 & 16.18$\pm$1.13 & 94.50$\pm$1.64 \\
    2459608.246 & 1500 & 218 & 14.04$\pm$1.28 & 94.49$\pm$1.64 \\
    2459609.221 & -- & 281 & 17.31$\pm$2.56 & 92.69$\pm$1.60 \\
    2459609.202$^{c}$ & 1000 & 227 & -- & -- \\
    2459610.220 & 1500 & 277 & 17.24$\pm$1.41 & 94.51$\pm$1.61  \\
    2459610.242$^{d}$ & 1000 & 249 & -- & -- \\
    2459612.185 & -- & 189 & 16.31$\pm$1.88 & 93.08$\pm$1.67 \\
    2459614.192 & -- & 170 & 16.53$\pm$1.39 & 92.68$\pm$1.79 \\
    2459622.193 & -- & 256 & 17.63$\pm$2.29 & 94.50$\pm$1.60 \\
    2459623.194 & -- & 203 & 18.41$\pm$2.63 & 92.22$\pm$1.76 \\
    \enddata
    \tablecomments{BJD 2459600 corresponds to UT Date 20220120.\newline
    a) All data taken from $\lambda=0.38-0.86 \mu m$. \newline b) Values reported are SNR per resolution element at 0.518 $\mu$m.\newline c, d) 2459609 and 2459610 both have 2 spectra. For each night, both spectra are combined to measure RV and \textit{v}sin(\textit{i}) signal.}
    \label{spectroscopic_obs}
    \vspace{-0.5cm}
\end{deluxetable}

HIP 5319 was observed during three different epochs in July 2020, September 2021, and January 2022 at the Subaru Telescope on Maunakea using SCExAO coupled with the CHARIS, MEC, and VAMPIRES instruments. During these epochs, the seeing conditions at the Subaru Telescope ranged between $\theta_{V}$=0\farcs{}4-0\farcs{}7. Observing conditions were photometric each night\footnote{The observing conditions during the January 2022 epoch were photometric, but due to instrument constraints there was no appropriate energy calibration of the MEC instrument, disallowing the measurement of a meaningful photometric data point.}. It was also observed for a fourth epoch in January 2022 at the W.M. Keck Observatory on Maunakea using the NIRC2 instrument coupled with the Keck Adaptive Optics system. The seeing during this epoch was $\theta_{V}$=0\farcs{}6. The observations from these runs are summarized in Table \ref{obslog_hip5319}.

All of the observations were taken with SCExAO using its ``vertical angle''/pupil-tracking mode which enables ADI \citep[][]{Marois2006}. Each set of data also used the Lyot coronagraph (0\farcs{}113 radius occulting mask) to suppress light from the primary star. The data in both epochs also utilized satellite spots for precise astrometric and spectrophotometric calibration \citep{Jovanovich2015a, Currie2018a}.

The MEC data in July 2020 was taken in $Y$ band (0.95-1.12 $\mu$m) with a spectral resolution $\mathcal{R}\sim$ 4.0 simultaneously with CHARIS broadband data. The CHARIS data in both epochs was taken in its low-resolution broadband mode covering $JHK$ passbands (1.16-2.37 $\mu$m) at $\mathcal{R}\sim$ 18. VAMPIRES data were taken at 750 nm concurrently with CHARIS in broadband mode in September 2021. In addition to the SCExAO observing mode allowing for ADI, the CHARIS spectral coverage enables SDI \citep[][]{Marois2000}. The NIRC2 data were taken in the $L_{p}$ filter ($\lambda_{\rm c}$ = 3.78 $\mu$m). Later in January 2022 more MEC data were taken covering $YJ$ bands (0.95-1.14 $\mu$m) with resolution $\mathcal{R}\sim$2.4. 

HIP 5319 was also observed for spectroscopic characterization of the primary during January and February of 2022. Spectra were obtained using the Network of Robotic Echelle Spectrographs (NRES) 1-m instrument operated by the Las Cumbres Observatory global telescope network \citep[LCOGT;][]{Brown2013} at the Wise Observatory in Mitzpe Ramon, Israel over the course of 9 nights from 20 January to 12 February 2022. These were taken using fiber-fed optical (0.38-0.86 $\mu$m) echelle spectrographs which have a spectral resolution of $\mathcal{R}\approx$50,000 and a SNR$>$200 for all but two of the spectra. The spectroscopic observations from the LCOGT NRES instrument are summarized in Table \ref{spectroscopic_obs}.

\subsubsection{CHARIS}
 We extracted CHARIS data cubes from the raw data using the standard CHARIS pipeline \citep{Brandt2017} to perform basic reduction steps --  image registration and spectrophotometric calibration. We did not obtain sky frames for sky subtraction.  For spectrophotometric calibration, we adopted a Kurucz stellar atmosphere model appropriate for an F5IV star.  HIP 5319 B is easily visible in the raw data for both CHARIS observations, but the September 2021 data suffered chronic PSF splitting due to low-wind effect, leaving us with only 8 exposures totaling just over 4 minutes of integration time.   The July 2020 data were stable: thus, we consider the September 2021 data only for astrometry and employ PSF subtraction to yield a high-quality spectrum for the July 2020 data.  
 
 To subtract the PSF in the July 2020 data,  we followed previous steps in \citet{Steiger2021}, using a full-frame implementation of reference star differential imaging (RDI) using the \textit{Karhunen-Loe`ve Image Projection} \citep[KLIP;][]{Soummer2012} algorithm as in \citet{Currie2019}.  Since the companion around HIP 5319 was easily visible, we adopted a conservative approach, truncating the KLIP basis set at one mode (KL = 1).   We corrected for minor throughput losses using KLIP forward-modeling as in \citet{Pueyo2016}.    
 %although results obtained with A-LOCI were similar \citep{Currie2012,Currie2015}.
 
\subsubsection{VAMPIRES}
\label{subsub:vampires}
 For VAMPIRES data, we subtracted dark frames and then aligned each sub-exposure within the 12.8 second data cubes, removing outliers. Subsequent steps used the general purpose high-contrast ADI broadband imaging pipeline from \citet{Currie2011}. To calibrate the VAMPIRES photometry an appropriate PHOENIX model stellar spectrum\footnote{\url{http://phoenix.astro.physik.uni-goettingen.de/}} \citep{PHOENIX2013} for an F5IV star was obtained and then normalized to the reported J band flux value for the HIP 5319 primary from the Two Micron All Sky Survey \citep[2MASS;][]{2MASS2006}. Once the model stellar spectrum had been calibrated, the flux density at 750 nm was found to be 13.18 Jy. For PSF subtraction, we found the best results with a full-frame implementation of ALOCI \citep{Currie2012,Currie2015}. Following \citet{Currie2018a}, we used forward-modeling to correct for throughput losses.
 
 %reference star differential imaging (RDI) using the \textit{Karhunen-Loe`ve Image Projection} \citep[KLIP;][]{Soummer2012} algorithm as in \citet{Currie2019a}, although results obtained with A-LOCI were similar \citep{Currie2012,Currie2015}.   For the 2015 NIRC2 data, we used a full-frame version of A-LOCI.  

\subsubsection{MEC}
\label{subsub:MEC}
$Y$ band images were created using the MKID Science Data Pipeline \citep{Steiger2022} to apply calibrations to the raw MEC data that include cold-, dead-, and hot-pixel masking, along with wavelength, astrometric, and spectrophotometric calibrations. There was no PSF subtraction performed for the data from MEC in this analysis.

The spectrophotometric calibration follows the treatment in \citet{Steiger2021} in which the flux from the elongated satellite spots in the image was measured using a ``racetrack'' aperture \citep{MillarBlanchaer2016} before being converted to the stellar flux behind the coronagraph using the relationship between satellite spot contrast and bandpass described in \citet{Currie2018a}. The stellar flux in the observation is then matched to the calibrated model spectrum from section \ref{subsub:vampires} to find a spectrophotometric solution, which is applied to the image to convert from counts per second units to units of flux density.

\subsubsection{NIRC2}
Our reduction steps followed ones outlined in \citet{Steiger2021}.  Briefly, we used a well-tested general purpose high-contrast ADI broadband imaging pipeline \citep{Currie2011} to perform basic processing, including sky subtraction, image registration, and photometric calibration.  To subtract the PSF, we used a full-frame implementation of reference star differential imaging (RDI) using the \textit{Karhunen-Loe`ve Image Projection} \citep[KLIP;][]{Soummer2012} algorithm as in \citet{Currie2019}.   The star BD+54 408 was used as a reference PSF.  Following \citet{Pueyo2016}, we used forward-modeling to correct for throughput losses.  
%\textcolor{red}{Quick section regarding Keck/NIRC2 data reduction.}

\subsubsection{NRES}
\label{subsub:LCOGT}
All spectra from the LCOGT 1-m NRES observations are automatically reduced using the BANZAI-NRES data reduction pipeline\footnote{Accessible at \url{https://github.com/lcogt/banzai-nres}}. After reduction, each spectrum was fit to the same F5IV star model stellar spectrum used in \ref{subsub:vampires} and \ref{subsub:MEC} and both RV and \textit{v}sin(\textit{i}) values for the primary were subsequently extracted using the H$\alpha$ and H$\beta$ spectral lines (nominally at $\lambda_{\alpha}=0.656\mu$m and $\lambda_{\beta}=0.486\mu$m) which are shown in Table \ref{spectroscopic_obs}.

The RV and \textit{v}sin(\textit{i}) values were calculated iteratively. For each spectra, an RV offset was fit via cross correlation with a PHOENIX model spectrum (the same that was used in sections \ref{subsub:vampires} and \ref{subsub:MEC} for VAMPIRES and MEC calibration) convolved to a first guess \textit{v}sin(\textit{i}) of 100 km/s. At that RV offset a \textit{v}sin(\textit{i}) is then calculated by minimizing $\chi^{2}$ between the model and NRES spectra, convolving over a grid of \textit{v}sin(\textit{i}) values between 50 and 150 km/s. This process is then iterated until the values for RV and \textit{v}sin(\textit{i}) converge, meaning that the scatter between the value of the most recent iteration and the previous is less than the formal error. The formal errors on the \textit{v}sin(\textit{i}) values are calculated using standard $\chi^{2}$ statistics. The formal error on the RV values are from the 1$\sigma$ confidence interval of the bootstrap probability density of the radial velocity.

\subsection{Detections}

\begin{figure*}
    \centering
    \includegraphics[width=0.9\textwidth]{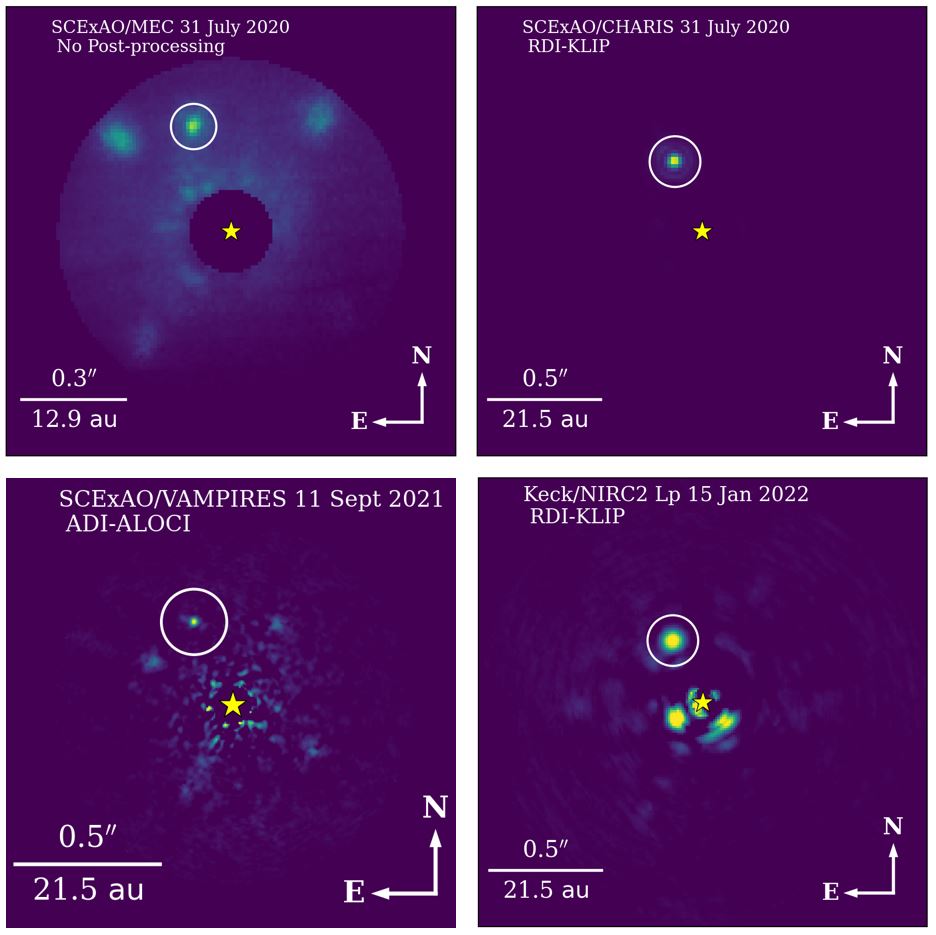}
       
    \caption{Detection of HIP 5319 B from SCExAO coupled with MEC, CHARIS, and VAMPIRES and Keck II Adaptive optics coupled with NIRC2. The MEC and VAMPIRES images retain some residual signal from satellite spots used for spectrophotometric and astrometric calibration. In MEC data, these spots appear with different brightnesses due to vignetting from the optics in MEC and dead pixels on the array, both of which have since been corrected. The NIRC2 image also retains some signal from the primary that was not removed by RDI-KLIP. The CHARIS data do retain some residual signal although the signal is so low that it cannot be seen without drastically lowering the maximum value of intensity in the image and saturating the PSF.
   % Total intensity image of HIP 5319 from SCExAO/MEC in \textit{y} band, with companion circled. The astrogrid, a set of 4 copies of the stellar PSF created by modulating the deformable mirror when the Lyot coronagraph is obscuring the central star, appears with different brightnesses due to vignetting from the optics in MEC and dead pixels on the array, both of which have since been corrected.
    }
    \label{fig:hip5319_obs}
\end{figure*}

Over all epochs the seeing conditions and data quality led to strong detections of the companion in each observing data set. To calculate flux density measurements in each channel, we performed aperture photometry sized to 1 $\lambda$/D.   The SNR was calculated in the standard fashion, replacing each pixel with the sum within an aperture, computing the robust standard deviation of these summed pixels as a function of angular separation and dividing by the stellar flux \citep{Marois2008b,Currie2011}.   Our spectrophotometric errors and SNR values consider finite-element corrections \citep{Mawet2014}.  SNR values range from 15.7 in the $Y$ band image from MEC to 763 in the broadband wavelength-collapsed CHARIS data taken in January 2022 and July 2020, respectively.  Following previous work, we use the IDL function \textrm{cntrd.pro} to estimate companion centroids: the error budget considers the intrinsic SNR of the detection, uncertainties in the plate scale and north position angle, and astrometric biases from processing \citep{Pueyo2016}.

In the July 2020 data, HIP 5319 B is located at [E,N]\arcsec{}$=$[0\farcs{}124, 0\farcs{}311]$\pm$[0\farcs{}004, 0\farcs{}004] and [0\farcs{}119, 0\farcs{}314]$\pm$[0\farcs{}010, 0\farcs{}010] in the CHARIS and MEC data, respectively. The errors in position take into account centroiding precision, the uncertainty in true north position angle, and pixel scale of each instrument following \citet{Currie2020}.

The September 2021 data from CHARIS and VAMPIRES show the companion at [E,N]\arcsec{}$=$[0\farcs{}133, 0\farcs{}287]$\pm$[0\farcs{}004, 0\farcs{}004] and [0\farcs{}132, 0\farcs{}287]$\pm$[0\farcs{}004, 0\farcs{}004]. The measurements taken by multiple instruments in both epochs are the same within error. The detections from each instrument are shown in Figure \ref{fig:hip5319_obs}. 

In January 2022, the NIRC2 and MEC data show the companion at [E,N]\arcsec{}$=$[0\farcs{}133, 0\farcs{}275]$\pm$[0\farcs{}003, 0\farcs{}003] and [0\farcs{}131, 0\farcs{}273]$\pm$[0\farcs{}010, 0\farcs{}010], where the MEC data were taken 4 days after the NIRC2 observations. 

Based on the proper motion of the primary between July 2020 and September 2021, a background star would have moved north-west by $\sim$ [-0\farcs{}23, 0\farcs{}03], which is inconsistent with the measured companion offset of [0\farcs{}009, -0\farcs{}024].

In standard Maunakea Observatory filters, the photometry for HIP 5319 B from the CHARIS broadband data is found to be $J=10.88\pm0.02$, $H=10.31\pm0.02$, and $K=10.07\pm0.03$ from the July 2020 data. These values are within 1$\sigma$ uncertainty for $H$ and $K$ band and 2$\sigma$ uncertainty for the measured $J$ band photometry points measured in September 2021. The MEC $Y$ band photometry is found to be $Y=11.3\pm0.1$, and VAMPIRES measured a flux density of 18.83 $\pm$ 0.83 mJy at 750 nm\footnote{For further discussion of the VAMPIRES photometry at 750 nm and its conversion to a pseudomagnitude see section \ref{sub:spectrum}.}.  Note that these measurements do not consider an absolute spectrophotometric uncertainty -- i.e. a multiplicative factor in flux density, additive in magnitude -- of 5\% due to uncertainties in the mapping between the deformable mirror modulation amplitude used to produce satellite spots and the resulting spot contrast at our fiducial wavelength of 1.55 $\mu m$ \citep{Currie2018b}. In the Keck II Telescope filters the photometry from the NIRC2 data is found to be $L_{p}=9.39\pm0.07$. The full summary of the HIP 5319 B detection significance, astrometry, and photometry is found in Table \ref{detections}.

\section{Analysis}

\begin{deluxetable*}{llllll}
    \tablewidth{0pt}
    \tabletypesize{\scriptsize}
    \tablecaption{HIP 5319~B Detection Significance, Astrometry, and Photometry}
    \tablehead{\colhead{UT Date} & \colhead{Instrument} & \colhead{Passband} & \colhead{SNR} & \colhead{[E,N](\arcsec{})} & {Photometry}}
    %& \colhead{$\sigma$[E,N]}}
    %Seeing (\arcsec{})} &{Passband$^{a}$} & \colhead{$\lambda$ ($\mu$m)$^{a}$} 
    %& \colhead{$t_{\rm exp}$} & \colhead{$N_{\rm exp}$} & \colhead{$\Delta$PA ($^{o}$)} & \colhead{Observing} \\
    %{} & {} & {} & {} & {} & {} & {} & {} & {} & \colhead{Strategy}  }
    \startdata
  %  \textbf{New Data}\\
    20200731 &  SCExAO/CHARIS & $JHK$ & 763 & [0.124, 0.311] $\pm$ [0.004, 0.004] & J = 10.88 $\pm$ 0.02 , H = 10.31 $\pm$ 0.02, K = 10.07 $\pm$ 0.03\\
    20200731  & SCExAO/MEC & $Y$ & 22.8 & [0.119, 0.314] $\pm$ [0.010, 0.010] & Y = 11.3 $\pm$ 0.1 \\
    20210911  &SCExAO/CHARIS &$JHK$ &48 & [0.133, 0.287] $\pm$ [0.004, 0.004] & J = 11.02 $\pm$ 0.06 , H = 10.38 $\pm$ 0.05, K = 10.09 $\pm$ 0.06\\
    20210911  &SCExAO/VAMPIRES & 750nm& 23 & [0.132, 0.287] $\pm$ [0.004, 0.004]& 18.83 mJy $\pm$ 0.83 mJy \\
    20220115  & Keck/NIRC2 & L$_{p}$ & 16.1 & [0.133, 0.275] $\pm$ [0.003, 0.003] & L$_{p}$ = 9.39 $\pm$ 0.067\\
    20220119  & SCExAO/MEC & $YJ$ & 15.7 & [0.131, 0.273] $\pm$ [0.010, 0.010] & -- \\
    \enddata
    \tablecomments{There is no photometry point measured during the 20220119 SCExAO/MEC observation. The CHARIS photometry do not consider an additional 0.05 magnitude uncertainty drawn from the mapping between the deformable mirror modulation amplitude (used to produce satellite spots used for spectrophotometric calibration) and the resulting satellite spot contrast with respect to the star.}
    \label{detections}
    \vspace{0cm}
\end{deluxetable*}

\begin{deluxetable}{llllllllll}
    \tablewidth{\columnwidth}
    \tabletypesize{\small}
    \tablecaption{HIP 5319 B Orbit Fitting Results and Priors}
    \tablehead{\colhead{Parameter} & \colhead{Fitted Value} &  \colhead{Prior}}
    \startdata
 %   \textbf{New Data}\\
    $M_{pri}$ ($M_{\odot})$ & $1.397^{+0.050}_{-0.052}$ & Gaussian,  $1.4 \pm 0.05$\\
    $M_{sec}$ ($M_{\rm Jup})$ & $31^{+35}_{-11}$ & 1/$M_{sec}$ (log flat) \\
    Semimajor axis \textit{a} (au) & $18.6^{+10}_{-4.1}$ & 1/a (log flat) \\
    Eccentricity \textit{e} & $0.42^{+0.39}_{-0.29}$ & uniform \\
    Inclination \textit{i} ($^{\circ}$) & $69.4^{+5.6}_{-15}$ & sin \textit{i} (geometric)\\
    \enddata
    \tablecomments{Posterior distributions for the secondary mass and semimajor axis are both positively skewed and favor low mass, low separation distributions. The eccentricity is not well constrained using only 2 relative astrometry points and no RV data, though future astrometry for this target should serve to better constrain this value.}
    \label{orbit_params}
    \vspace{-1.0cm}
\end{deluxetable}

\subsection{Characterization of HIP 5319 A as a Single Star}
\label{sub:primary_single_star}

\begin{figure*}
    \centering
    \includegraphics[width=1.05\columnwidth]{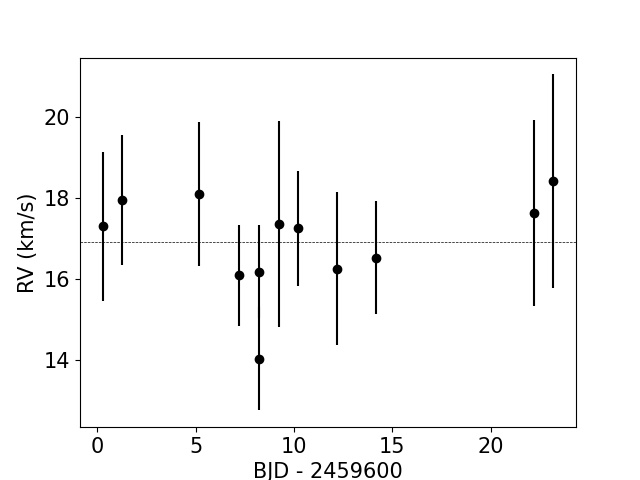}
    \includegraphics[width=1.05\columnwidth]{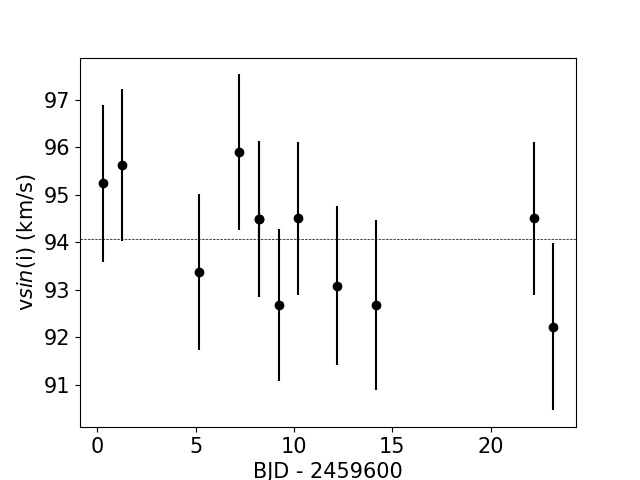} \\
    \includegraphics[width=1.05\columnwidth]{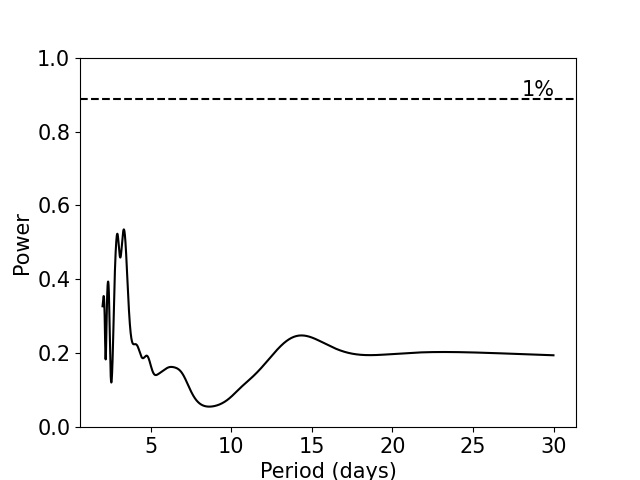}
    \includegraphics[width=1.05\columnwidth]{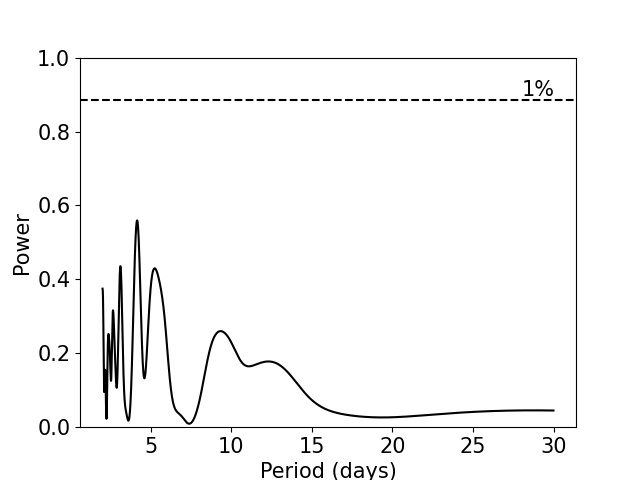}
    \caption{(Top) Radial velocity (left) and \textit{v}sin(\textit{i}) (right) values measured for HIP 5319A. The dotted lines in each panel are the best fit constant velocity to the data, where RV=16.71 km/s and \textit{v}sin(\textit{i})=94.21 km/s. Neither metric shows either significant variation in time or obvious periodicity. (Bottom) Periodograms of the residuals from the radial velocity (left) and \textit{v}sin(\textit{i}) (right) values. The residuals for each metric are calculated by taking the measured data and subtracting the best fit constant velocity. The false alarm probability of 1$\%$, calculated using bootstrap randomization, is shown by the dashed lines.}
    \label{fig:spectroscopic_timeseries}
\end{figure*}

\begin{figure}
    \centering
    \includegraphics[width=\columnwidth]{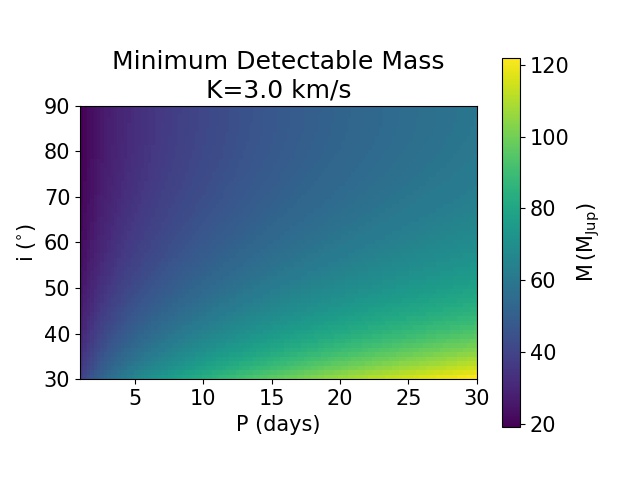} \\
    \includegraphics[width=.95\columnwidth]{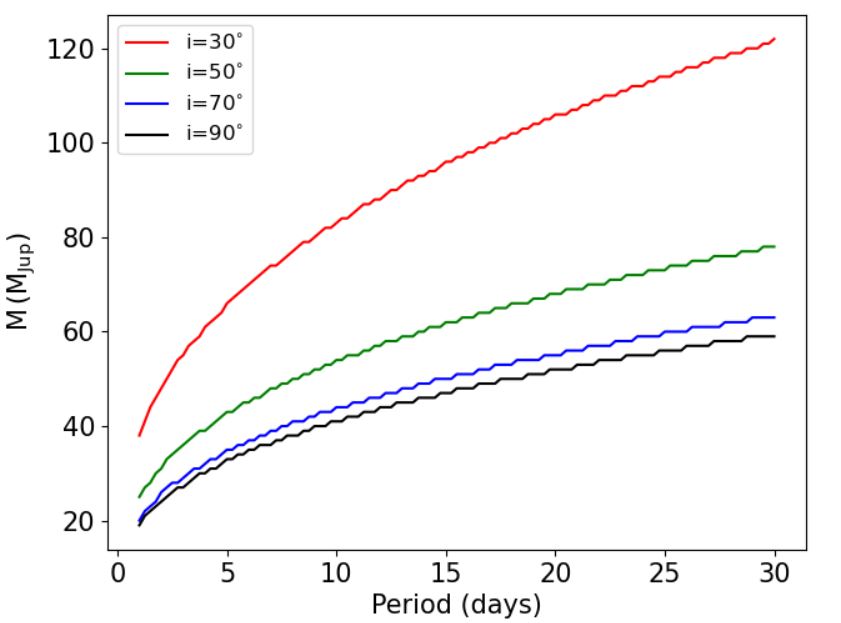}
    \caption{(Top) Minimum detectable binary companion mass for various periods ($P$) and inclinations ($i$). For a given combination of period and inclination, the reported `minimum detectable mass' can be found, which corresponds to the lowest mass a companion would have that would result in a semiamplitude $K>3$ km/s. Any companions less than that mass would be undetectable in the spectroscopic data and any companions that were more massive would have been detected. (Bottom) Minimum detectable mass as a function of period for selected inclinations.}
    \label{fig:minimum_detectable_mass}
\end{figure}

Before the properties of the companion can be determined it is first necessary to identify whether the primary is a binary or single star. Using the RV and \textit{v}sin(\textit{i}) values we look for periodic variations in time to help infer the presence of a companion or lack thereof. The top panels in Figure \ref{fig:spectroscopic_timeseries} show the measured values of each quantity and the best fit to a constant velocity.

The search for \textit{v}sin(\textit{i}) is motivated due to large scatter in this quantity's previously reported values in the literature, which range from 35 km/s \citep[][]{Nordstrom2004} to 125 km/s \citep[][]{Danziger1972} at the low and high ends, respectively. In the collection of stars discussed in \citet{Fleming1989} HIP 5319 has the greatest uncertainty on its \textit{v}sin(\textit{i}) value, nearly double the next highest uncertainty and almost 1/3 of its reported rotation rate. This wide scatter in reported rotation rates along with the high uncertainties reported on these measurements led us to consider whether there may be a binary companion where both objects contribute to the spectrum whose individual signals have not been teased out. Since we can obtain \textit{v}sin(\textit{i}) from the NRES spectra we use this opportunity to search for any signal in the data which may indicate the presence of a second, unseen companion contaminating the signal from the primary star.

The bottom panels in figure \ref{fig:spectroscopic_timeseries} show periodograms of the residuals from the RV and \textit{v}sin(\textit{i}) data. The peak values of each periodogram are 0.559 and 0.535, respectively. Assuming there is no periodic signal in the data, this means that a peak this high or higher will be seen 79.6$\%$ of the time in the RV data and 67.6$\%$ of the time in the \textit{v}sin(\textit{i}) data. Also shown are the required peak heights to attain a 1$\%$ false alarm probability for each measurement. For the radial velocity data a peak would have to have a power of 0.888 to attain a false alarm probability below 1$\%$, while the \textit{v}sin(\textit{i}) peak would need to have a power of 0.894 to meet the same criterion. The height of the 2 peaks from the periodograms combined with the high peak values needed to attain a 1$\%$ false alarm probability demonstrate that there is no obvious periodic signal, meaning the time series RV and \textit{v}sin(\textit{i}) data are not consistent with oscillatory behavior caused by a close-in companion.

Both sets of measurements are consistent with constant values to within 1 standard deviation except for a single point: the radial velocity measured from the second spectrum on BJD 2459608. In both cases we see that we would  be sensitive to any periodic signal with a semi-amplitude $K\gtrsim3$ km/s, while any signal that has $K\lesssim3$ km/s may still be hidden within the measurement error.

Using equation \ref{eqn:semiamplitude} - which relates the semi-amplitude $K$ to the orbital period $P$ of a companion of mass $M_{2}$ around a host of mass $M_{1}$ with eccentricity and inclination $e$ and $i$ - it is possible to estimate the detectable companion mass for a given set of $P$, $i$, and $e$ values.

\begin{equation}
\label{eqn:semiamplitude}
    K = \bigg(\frac{2\pi G}{P}\bigg)^{1/3}\frac{M_{2}\text{sin}(i)}{(M_{2}+M_{1})^{2/3}}\frac{1}{\sqrt{1-e^{2}}}
\end{equation}

For this estimation $K_{max}=3$ km/s and $e$ is assumed to be equal to 0. We then vary $P$ and $i$ and calculate the smallest mass that would generate an RV semi-amplitude $K>K_{max}$ for each ($P$, $i$) combination. The results of this are shown in Figure \ref{fig:minimum_detectable_mass} for 2$\leq P\leq$30 days and 30$^{\circ}\leq i\leq90^{\circ}$. 

The choice to restrict this analysis to periods between 2 and 30 days is due to the cadence of observations and the duration of the survey. A companion with a shorter period may still have been detectable although without being able to accurately measure the period. We would not have enough data to detect a companion with a period P$\gtrsim$30 days since there would be insufficient time to see periodicity in the signal; however, our data do cover the range of expected periods for an RS CVn system (P$\lesssim$14 days). With regards to the inclination the analysis is not performed below 30$^{\circ}$ due to the difficulty of detecting companions in RV signals for near face on orbits. The original claim was of this star as a spectroscopic binary, meaning that the system would not have been face on.

At the extreme values of the analysis we find that a binary companion with $P=$2 days and $i=90^{\circ}$ would be detectable if it had a mass greater than 24 $M_{\rm J}$ whereas for a companion with $P=$30 days and $i=30^{\circ}$ the minimum mass that would be detectable via an RV signal would be 122 $M_{\rm J}$. This tells us that in the spectroscopic data taken on this star we would have seen the signature for a binary companion above 122 $M_{\rm J}$ at worst and 24 $M_{\rm J}$ at best.

Further spectroscopic data taken at higher precision and over longer times will aid in ruling out potential lower mass and longer period binary companions, but current data suggest there is no companion with mass greater than 122 $M_{\rm J}$ with a duration less than 30 days, which is sufficient to refute previous evidence of this star being a spectroscopic binary.

\subsection{Non-detection of Ca HK Emission}
\label{sub:ca_hk}

\begin{figure*}
    \centering
    \includegraphics[width=\columnwidth]{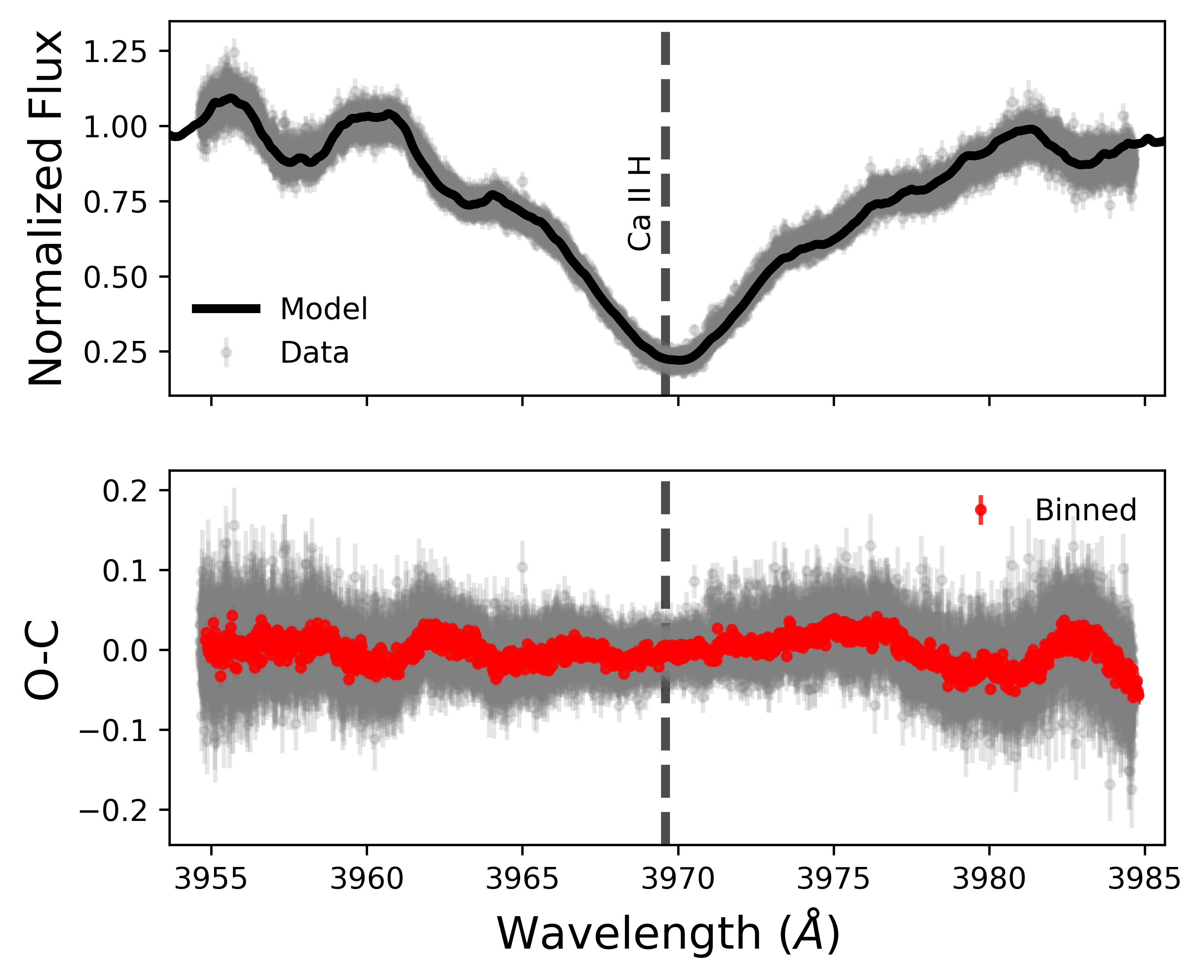}
    \includegraphics[width=\columnwidth]{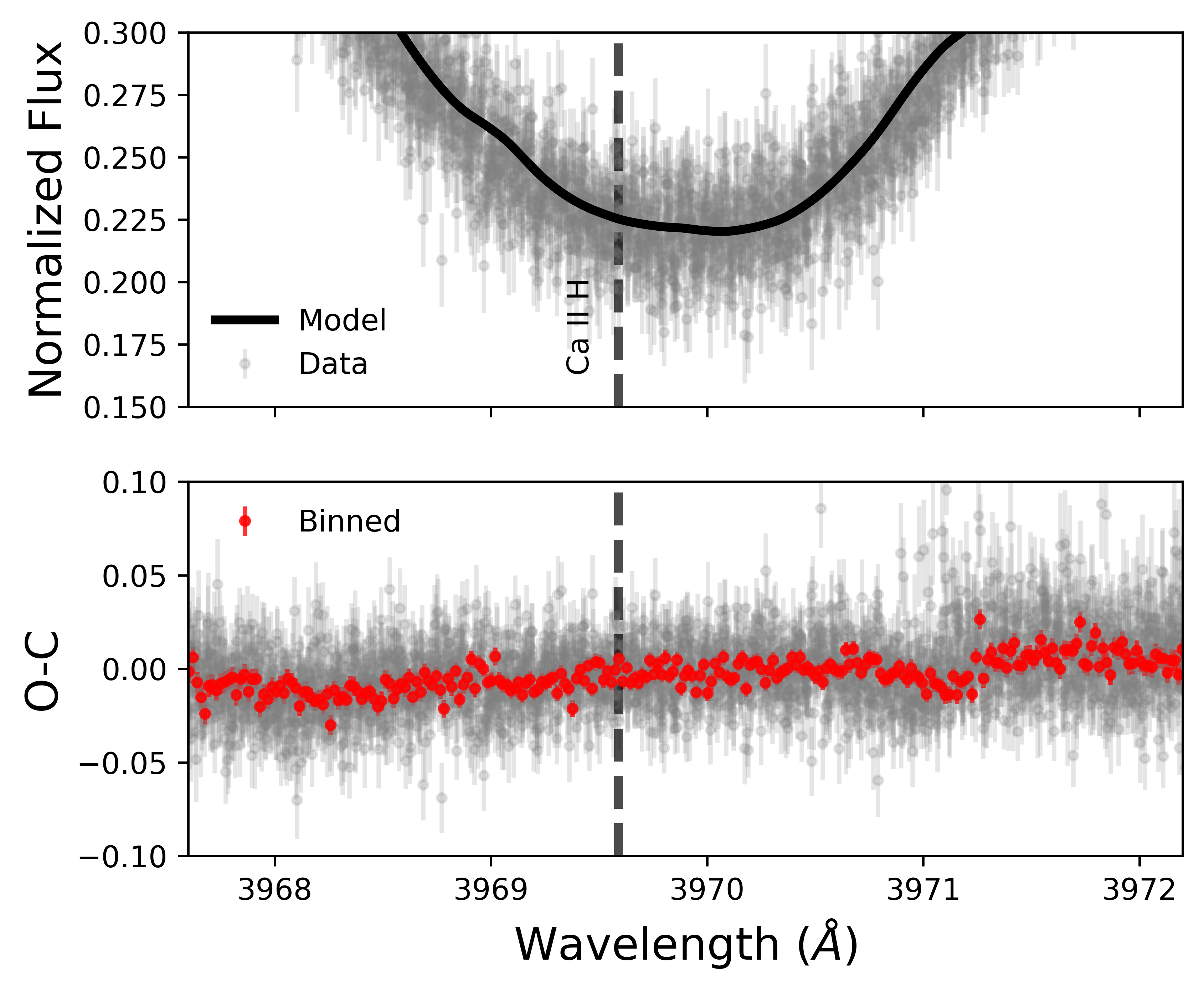}
    \caption{Data from LCOGT NRES spectra of the primary star HIP 5319 A compared to a PHOENIX model spectrum for an F5IV star surrounding the Ca II H line at its vacuum wavelength $\lambda=3969.5\AA$. The model spectrum has been broadened by 100 km/s to match the best-fit \textit{v}sin(\textit{i}) value for the Ca II H line from the LCOGT spectra. (Left top) Model spectra plotted over data from the 12 LCOGT spectra between $\lambda=3955-3985\AA$. (Left bottom) The O$-$C (Observed$-$Calculated) plot showing the residuals between measured data and model. Grey points are the residuals from each of the 12 spectra, while the red points are rebinned to the original NRES spectral resolution. (Right) The same data and residuals between $\lambda=3968-3972\AA$. In both cases it can be clearly seen that there is no excess flux beyond the 1\% level in the spectrum at any point near the Ca H line.}
    \label{fig:ca_hk_nondetection}
\end{figure*}

%The second facet of the misidentification of HIP 5319 is its reported high chromospheric activity.
We also reassess evidence that HIP 5319 has a high chromospheric activity. \citet{BoroSaikia2018} previously claim to have measured a value of log($R\prime_{HK}$)=-4.016. 
The methodology behind this claim was to measure the surface flux $R_{HK}$ by co-adding all available spectra for the target into a template spectra that was then normalized to a PHOENIX model atmosphere in order to convert to absolute flux units. The photospheric flux contribution $R_{phot}=F_{phot}/\sigma T_{eff}^{4}$ was then subtracted from the integrated flux of the Ca II H and K line cores from the PHOENIX model atmosphere. The excess that was seen after this subtraction interpreted as being from emission at the H and K lines.

By comparing the high resolution LCOGT spectra (section \ref{subsub:LCOGT}, Table \ref{spectroscopic_obs}) and the model PHOENIX spectrum for an F5IV star used for photometric calibration (sections \ref{subsub:vampires}, \ref{subsub:MEC}) we find no evidence to support the claim of any excess flux around the Ca II H or K lines beyond the 1$\%$ level. 

To compare the difference between the model PHOENIX spectrum and LCOGT spectra, each nightly spectrum was individually normalized using a scale factor, slope, and offset. Figure \ref{fig:ca_hk_nondetection} shows the result of this comparison for the Ca II H line at $\lambda=3969.5\AA$.

The top panels in Figure \ref{fig:ca_hk_nondetection} show the data from all of the spectra in Table \ref{spectroscopic_obs} compared to the model PHOENIX spectrum, while the bottom panels show the residuals between the data and model. The residuals from each spectra compared to the model are shown as grey points, while the red points show the residuals when the data are rebinned to the original NRES R$\sim$50,000. This rebinning was performed because each spectra that makes up the combined dataset (made of 12 individual spectra) samples slightly different rest-frame wavelengths because of the evolving barycenter velocity over the 23 days where spectra were collected. This means that there is roughly 12 times as much data since the same wavelengths are not sampled multiple times. By rebinning to the original NRES resolution this has the effect of demonstrating what a single spectra would look like for $\sim$12 times as much observation time as one of the individual spectra on its own.

The data collected in the 14 observations match the model without any significant deviation around the cores of the Ca II HK lines. Along with the non-detection of an time-varying signal in the RV and \textit{v}sin(\textit{i}) data this refutes the evidence that the primary is an RS CVn binary which is expected to have high chromospheric activity and a period below 14 days, meaning it is likely a single star. This is in good agreement with the report of HIP 5319A from the Gaia Early Data Release 3 \citep[Gaia eDR3;][]{GaiaEDR3} as being well fit by a 5-parameter single star solution whose Renormalized Unit Weight Error (RUWE) is 1.01, which effectively rules out stellar-mass companions greater than $\sim 0.4M_{\odot}$ and a period between 1 and 10 days.

\subsection{Spectrum of HIP 5319 B}
\label{sub:spectrum}

\begin{figure}
    \centering
    \includegraphics[width=\columnwidth]{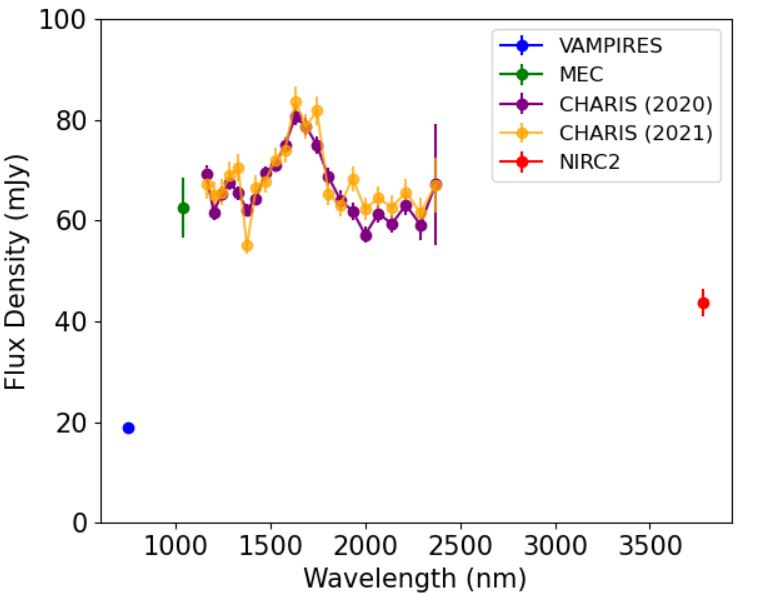}
    \caption{Combined SCExAO/CHARIS spectra, SCExAO/MEC photometry, SCExAO/VAMPIRES and Keck/NIRC2 photometry of the low mass companion HIP 5319 B taken on July 31, 2020 (CHARIS and MEC), September 11, 2021 (CHARIS and VAMPIRES), and January 15, 2022 (NIRC2) at the Subaru and Keck II telescopes. The reddest CHARIS channel has substantially higher uncertainty in our spectrophotometric calibration, because we did not obtain sky frames.}
    \label{fig:MEC+CHARIS_spectrum}
\end{figure}

\begin{figure}
    \centering
    \includegraphics[width=\columnwidth]{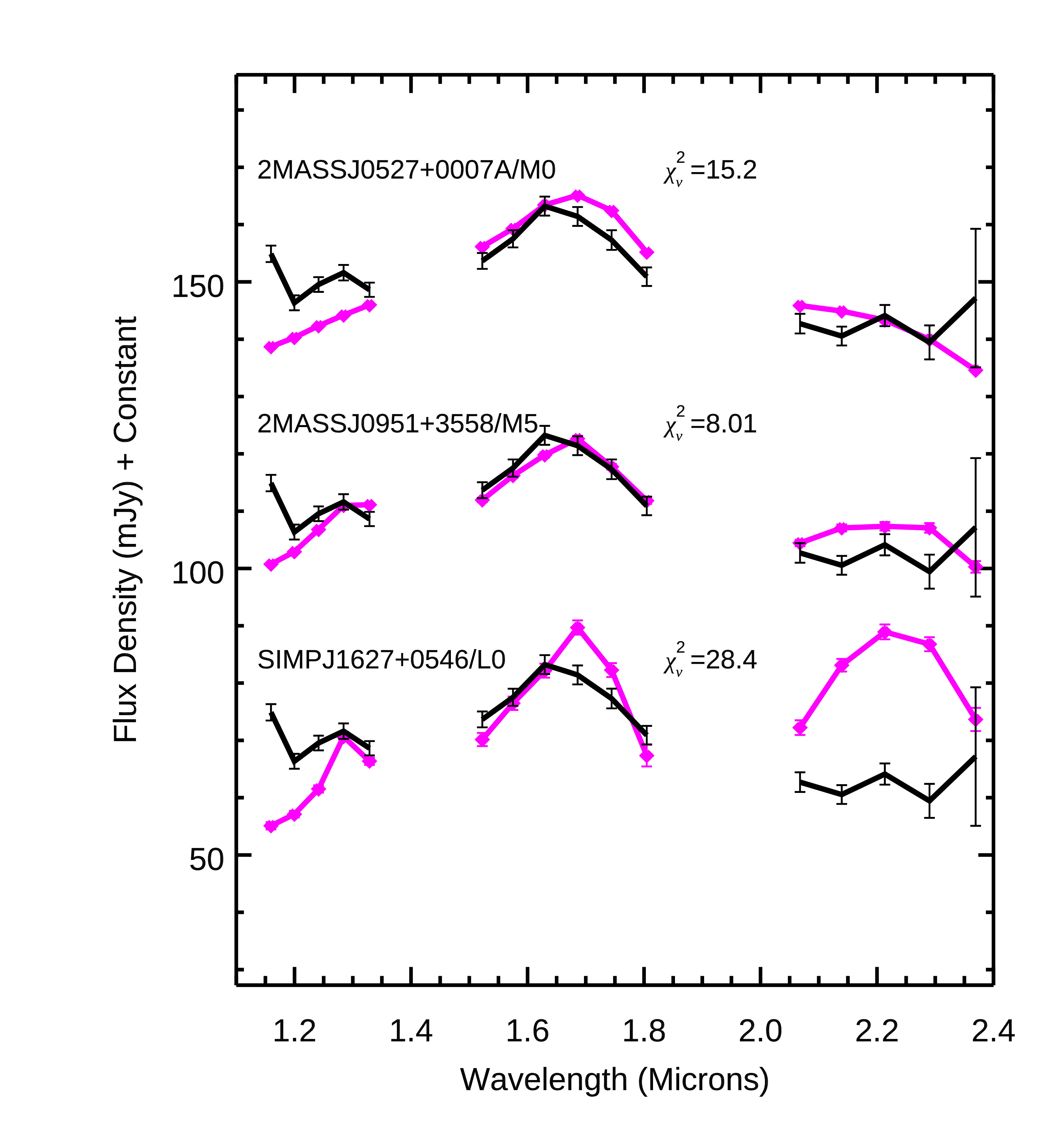}
    \caption{The CHARIS HIP 5319 B spectrum (black) compared to those of field brown dwarfs (magenta) with spectral types M0, M5, and L0 from the Montreal Spectral Library binned to CHARIS's resolution.}
    \label{fig:Hip5319_B_with_model_spectra}
\end{figure}

\begin{figure*}
    \centering
    \includegraphics[width=\columnwidth]{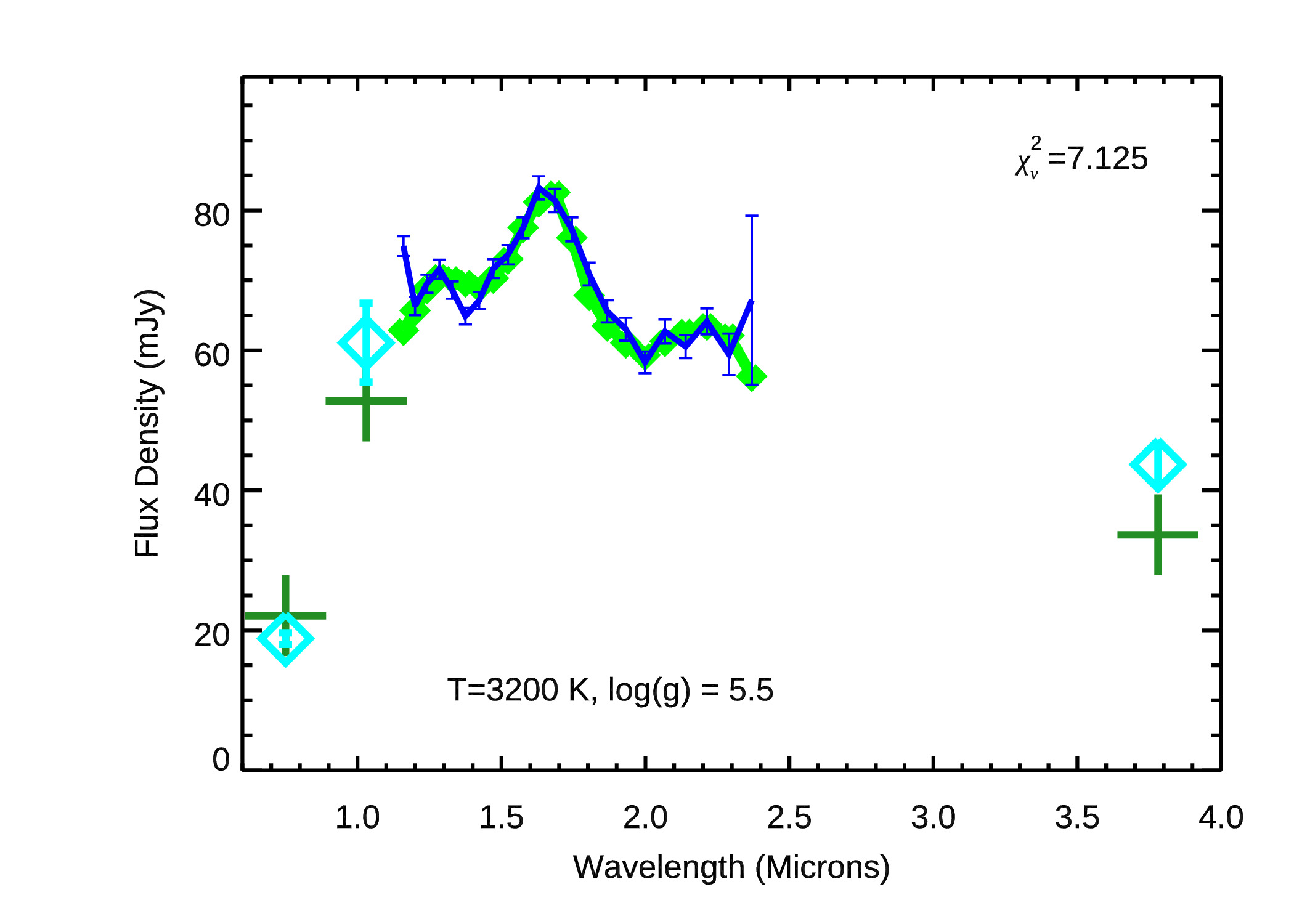}
    \includegraphics[width=\columnwidth]{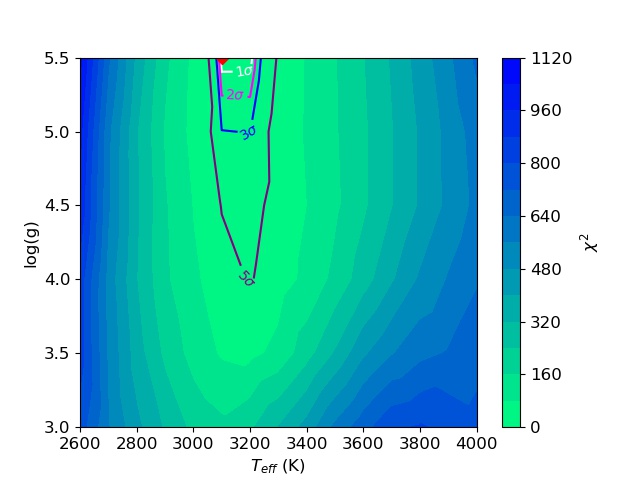}
    \caption{(Left) BT-Settl model for solar metallicity  with T=3200 K and log(\textit{g})=5.5.  CHARIS spectra is shown in dark blue, VAMPIRES, MEC and NIRC2 photometry in cyan compared to the model-predicted CHARIS spectrophotometry in light green, and predicted VAMPIRES/MEC/NIRC2 photometry (dark green crosses). Although the SNR of the spectrum is quite high, the spectral covariance in the CHARIS data is also high, leading to a large value of $\chi^{2}$. (Right) Corresponding contour plots for $\chi^{2}$ as a function of temperature and surface gravity. The best-fit solution is shown with a red diamond while the 1$\sigma$, 2$\sigma$, 3$\sigma$, and 5$\sigma$ contours are shown in white, magenta, blue, and purple, respectively.}
    \label{fig:btsettlfit}
\end{figure*}

Figure \ref{fig:MEC+CHARIS_spectrum} shows the 2020 and 2021 CHARIS spectra (whose data can be found in Table \ref{CHARIS_spectra}) as well as MEC\footnote{Although the MEC data has a median spectral resolution $\mathcal{R}\sim$ 4.0, we bin our spectral data to a single $Y$ band photometry point for comparison with the standard photometric band.}, VAMPIRES, and NIRC2 photometric points. The MEC photometry and CHARIS spectra are flat in $F_{\rm \nu }$ units except for a broad peak in $H$ band. Formally, the SNR of HIP 5319 B in each spectral channel is extremely high (SNR $>$ 77).   Outside of the $H$-band peak, consecutive wavelength channels show a ``wavy" pattern, which may indicate the impact of spectrally correlated noise (see below).   The two CHARIS spectra show broad agreement: due to the higher SNR for the 2020 epoch spectrum, we focus on it for subsequent analysis.

\begin{deluxetable*}{cllcrllcr}
    \tablewidth{0pt}
    \tabletypesize{\scriptsize}
    \tablecaption{HIP 5319 B Spectra}
    \tablehead{\colhead{} & \vline & \colhead{} & \colhead{31 July 2020} & \colhead{} & \vline & \colhead{} & \colhead{11 September 2021} & \colhead{} \\ \colhead{Wavelength ($\mu$m)} & \vline & \colhead{$F_{\nu}$ (mJy)} & \colhead{$\sigma F_{\nu}$ (mJy)} & \colhead{SNR} & \vline & \colhead{$F_{\nu}$ (mJy)} & \colhead{$\sigma F_{\nu}$ (mJy)} & \colhead{SNR}}
    %& \colhead{$\sigma$[E,N]}}
    %Seeing (\arcsec{})} &{Passband$^{a}$} & \colhead{$\lambda$ ($\mu$m)$^{a}$} 
    %& \colhead{$t_{\rm exp}$} & \colhead{$N_{\rm exp}$} & \colhead{$\Delta$PA ($^{o}$)} & \colhead{Observing} \\
    %{} & {} & {} & {} & {} & {} & {} & {} & {} & \colhead{Strategy}  }
    \startdata
  %  \textbf{New Data}\\
     1.160 & \vline & 69.197 & 1.720  & 61.8  & \vline & 67.313 & 3.076 &  52.7 \\
     1.200 & \vline & 61.625 & 1.612  & 56.7  & \vline & 65.107 & 2.733 &  55.5 \\
     1.241 & \vline & 65.347 & 1.490  & 72.5  & \vline & 65.612 & 2.589 &  55.9 \\
     1.284 & \vline & 67.608 & 1.490  & 84.7  & \vline & 68.917 & 2.856 &  47.0 \\
     1.329 & \vline & 65.402 & 1.407  & 82.0  & \vline & 70.522 & 2.650 &  58.4 \\
     1.375 & \vline & 62.158 & 1.237  & 111.0 & \vline & 55.005 & 1.672 &  78.1 \\
     1.422 & \vline & 64.376 & 1.308  & 106.7 & \vline & 66.510 & 2.453 &  77.4 \\
     1.471 & \vline & 69.387 & 1.408  & 118.3 & \vline & 67.856 & 2.367 &  93.6 \\
     1.522 & \vline & 71.086 & 1.419  & 141.1 & \vline & 71.960 & 2.449 &  85.6 \\
     1.575 & \vline & 74.866 & 1.518  & 150.9 & \vline & 74.057 & 2.521 &  71.3 \\
     1.630 & \vline & 80.553 & 1.703  & 130.8 & \vline & 83.695 & 2.936 &  76.6 \\
     1.686 & \vline & 78.667 & 1.716  & 116.0 & \vline & 78.616 & 2.461 &  81.6 \\
     1.744 & \vline & 74.888 & 1.802  & 101.4 & \vline & 81.785 & 2.802 &  78.7 \\
     1.805 & \vline & 68.638 & 1.745  & 83.4  & \vline & 65.169 & 2.220 &  66.7 \\
     1.867 & \vline & 64.104 & 1.793  & 67.1  & \vline & 63.105 & 2.272 &  74.0 \\
     1.932 & \vline & 61.858 & 1.793  & 71.5  & \vline & 68.272 & 2.493 &  117.6 \\
     1.999 & \vline & 57.205 & 1.672  & 77.4  & \vline & 62.256 & 2.167 &  94.3 \\
     2.068 & \vline & 61.378 & 1.770  & 102.2 & \vline & 64.435 & 2.329 &  89.4 \\
     2.139 & \vline & 59.341 & 1.688  & 109.1 & \vline & 62.607 & 2.294 &  87.4 \\
     2.213 & \vline & 63.070 & 1.926  & 93.5  & \vline & 65.636 & 2.625 &  83.0 \\
     2.290 & \vline & 59.136 & 3.066  & 64.8  & \vline & 61.504 & 2.955 &  87.8 \\
     2.369 & \vline & 67.170 & 12.091 & 68.5  & \vline & 67.050 & 5.416 & 53.3
    \enddata
    \tablecomments{Throughput-corrected HIP 5319 B spectra extracted from July 2020 and September 2021 CHARIS data.}
    \label{CHARIS_spectra}
    \vspace{0cm}
\end{deluxetable*}

HIP 5319 B's broadband near-IR colors ($J$-$H$ $\sim$ 0.57 $\pm$ 0.03; $H$-$K_{\rm s}$ $\sim$ 0.24 $\pm$ 0.03) resemble those of early to mid M dwarfs \citep{Pecaut2013}.   HIP 5319 B is substantially fainter than the primary in the VAMPIRES 750 nm data ($\Delta m$ $\sim$ 7.110).  The VAMPIRES filter does not correspond to any standard photometric bandpass with a published zeropoint flux density but lies between the Johnson-Cousins $R$ and $I$ bands.   Adopting again the standard colors from \citet{Pecaut2013} and $R$ band optical photometry for the primary from the \textit{Simbad} database, we estimate a pseudomagnitude of $\approx$ 13 at 750 nm.

We compare HIP 5319 B's CHARIS spectrum with other low-mass objects in the Montreal Spectral Library\footnote{\url{https://jgagneastro.com/the-montreal-spectral-library/}} \citep[e.g.][]{Gagne2015}. Only the CHARIS spectrum was used because the wavelength range for the Montreal Spectral Library covers \textit{JHK}, but is rather non-uniform otherwise. Following the methods described in \citet{GrecoBrandt2016}, we find that the CHARIS spectrum shows noise that is highly spatially and spectrally correlated (A$_{\rm \rho}$ $\sim$ 0.69, A$_{\rm \lambda}$ $\sim$ 0.22).   
 HIP 5319 B is best matched by an M3--M7 dwarf: earlier M dwarfs and L dwarfs fail to reproduce the CHARIS spectra, especially in the $J$ and $K$ bands (see Figure \ref{fig:Hip5319_B_with_model_spectra}). 

%considering the impact of spatially and spectrally correlated noise \citep{GrecoBrandt2016}\footnote{We do not also compare the MEC or NIRC2 photometry due to sparse coverage of the library outside of the $JHK$ passbands}.   The CHARIS data reveal highly correlated errors (Figure \ref{fig:spectrum}, right panel).  The spectral covariance at HD 109427 B's location includes substantial off-diagonal terms, especially for spatially-correlated noise (A$_{\rm \rho}$ $\sim$ 0.71) and (to a lesser extent) residuals speckles well correlated as a function of wavelength (A$_{\rm \lambda}$ $\sim$ 0.16). 

Following similar analysis in \citet{Steiger2021}, we compared the MEC, VAMPIRES, and NIRC2 photometry and CHARIS spectrum to the BT-Settl atmosphere models \citep{Allard2012} with the \citet{Asplund2009} abundances and solar metallicities.
%downloaded from the Theoretical Spectra Web Server\footnote{\url{http://svo2.cab.inta-csic.es/theory/newov2/}}.    The grid covers temperatures of 2500--4000 $K$, surface gravities of log(g) = 3.5--5.5, and metallicities of [Fe/H] = -1 to 0.5.   
%Following \citet{Currie2018b}, 
We focus only on the CHARIS channels unaffected by telluric absorption and also remove the first CHARIS channel, whose high flux density is not reproduced in any empirical spectrum in the Montreal Library.    We define the fit quality for the $kth$ model using the $\chi^{2}$ statistic, considering the spectral covariance.
%\begin{equation}
%    \chi^{2} = R_{k}^{T}C^{-1}R_{k} + \sum_{i}(f_{phot,i}-\alpha_{k}~F_{phot,ik})^{2}/\sigma_{phot,i}^{2}.
%\end{equation}
%Here, the vector $R_{k}$ is the difference between measured and predicted CHARIS data points ($f_{spec}-\alpha_{k}F_{spec}$) and $C$ is the covariance for the CHARIS spectra.  The vectors $f_{phot,i}$, $F_{phot,ik}$, and $\sigma_{phot,i}$ are measured photometry, model predicted photometry, and photometric uncertainty; $\alpha_{k}$ is the scaling factor for the model that minimizes $\chi^{2}$ \citep[see also][]{DeRosa2016}.

Figure \ref{fig:btsettlfit} shows the best-fit solar metallicity model and associated $\chi^{2}$ contours. An atmosphere with a temperature of $T_{\rm eff}$ = 3100--3200 K and a high gravity (log(g) = 5.5) fits the data the best\footnote{Fits at 3100 $K$ and 3200 $K$ are almost numerically equivalent.}, although the family of solutions drawn from high gravity models (log(g) = 5--5.5) at 3100 K and 3300 K and those at 3200 K and a lower gravity of log(g) = 4--4.5 fall within $5\sigma$ of the best-fit model. The radii that minimize $\chi^{2}$ are 3.25--3.62 $R_{\rm J}$, yielding a luminosity of log(L/L$_{\rm \odot}$) = -1.94 $\pm$ 0.04. The best-fitting atmospheric models (log(g)=5.5, $R_{sec}=3.4-3.59R_{J}$) correspond to a companion whose mass is $\sim448-1675M_{\rm Jup}$, or $0.427-1.60M_{\odot}$. Some of these values would be significantly higher than those for a typical M3-M7 star \citep{Pecaut2013}: potentially greater than the mass of the primary itself.   However, 5-$\sigma$ confidence interval containing lower gravity solutions implies masses down to 44 $M_{\rm J}$ and includes a wider range of radii (3.25--3.62 $R_{\rm J}$).   Thus, while the temperature of HIP 5319 B is well constrained to 3100-3300 $K$, the companion's poorly constrained surface gravity results in poor mass limits.
%the implied masses range is poorly between 38 $M_{\rm J}$ and 88 $M_{\rm J}$. 
%The 1-$\sigma$ contour for temperature and gravity is narrowly defined about this peak for both metallicities: $T_{\rm eff}$ = 3100--3300 $K$ and log(g) = 5.25--5.5.   At the 2-$\sigma$ level, the best-fit temperature and gravity ranges widen to 3000--3400 $K$ and log(g) = 5--5.5.   The radii that minimize $\chi^{2}$ are $\sim$2.1--2.6 Jupiter radii.
%\textcolor{red}{stuff about atmospheric models}

Using isochrones from \citet{Baraffe2015} we find that using the age estimate of 1.07-1.23 Gyr from the Padova and BASTI models in \citep[][]{Casagrande2011} and adopting the luminosity of log(L/L$_{\rm \odot}$=-1.94$\pm$0.04) from the atmospheric models we estimate the mass of the secondary would fall between roughly 0.3-0.35$M_{\odot}$. Considering the \textit{widest} possible range of ages of 8 Myr to 2 Gyr (the lowest end predicted by \citet{StanfordMoore2020} and the highest predicted by \citet{Holmberg2009}) we find that the range of masses extends from ~40$M_{\rm Jup}$ to 0.35$M_{\odot}$. Both possible ranges include typical masses of M dwarfs from \citet{Pecaut2013}, while the low end of the range suggest masses down to 40$M_{\rm Jup}$, which does not disagree with either the dynamical mass (see section \ref{sub:dynamicalmass}) or the mass estimated from the atmospheric models above.

\subsection{Orbit and Dynamical Mass}
\label{sub:dynamicalmass}

\begin{figure*}
    \centering
    \includegraphics[width=\textwidth]{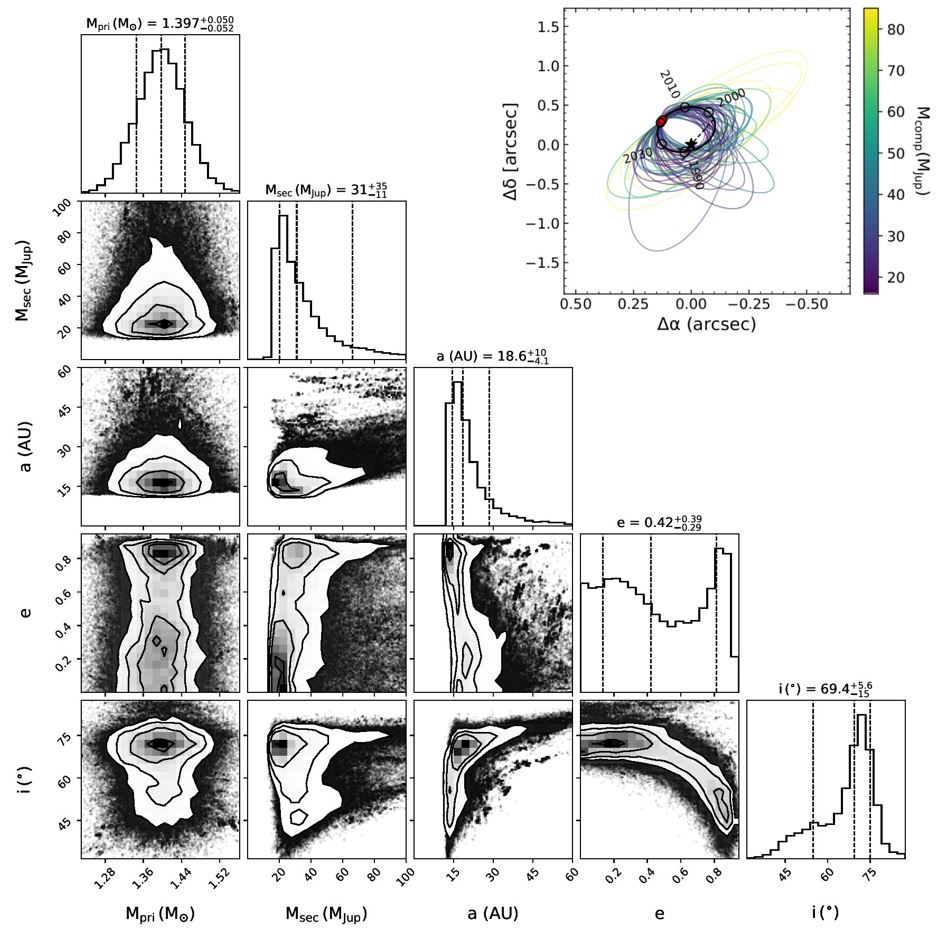}
    \caption{Corner plot showing the posterior distributions of selected orbital parameters using a log-normal (1/M) prior for the mass of the secondary companion. The orbit fits used \textit{Hipparcos} and \textit{Gaia} (HGCA) absolute astrometry and relative astrometry from SCExAO/CHARIS and MEC data. The inset in the figure shows the best fit orbit (black) with 50 random orbits drawn from the MCMC fits color coded by the mass of HIP 5319 B. The red-colored points in the orbit represent relative astrometry points from the 3 epochs where data were taken, and the unfilled circles show the predicted location of the companion at different past and future epochs. The companion is orbiting counterclockwise.}
    \label{fig:cornerplot}
\end{figure*}

We used the open-source code \texttt{orvara} \citep{orvara2021} to fit for the mass and orbit of HIP 5319 B. \texttt{orvara} uses a combination of radial velocity (RV) absolute astrometry of the primary, and relative astrometry of the low-mass companion to measure orbital parameters even when the observations of the companion only cover small fractions of an orbit. 

\subsubsection{Results Using a 1/$M_{\rm p}$ Prior for Companion Mass}

For this companion, we used HGCA absolute astrometry measurements for the star and three epochs of relative astrometry from CHARIS, MEC, and NIRC2. There is no archival RV data for this target and so it is not included in the \texttt{orvara} fits. A Gaussian prior of 1.4$\pm$0.05$M_{\odot}$ was chosen based on literature values for the primary mass \citep{Casagrande2011}, while a log-flat (1/M) prior was chosen for the mass of HIP 5319 B, which is the default used by \texttt{orvara}. This choice is motivated by the shape of the initial mass function for low-mass objects and for planets, which says that low-mass objects are expected to occur more frequently than high-mass ones \citep{Chabrier2003,Nielsen2019}.
%Without existing literature estimates on the mass of the compaion we therefore adopt a typical orbital prior for a companion: $p(M_{2}) \propto 1/M_{2}$ \citep[][]{Kuzuhara2022}}

Figure \ref{fig:cornerplot} shows the posterior distributions for the primary and secondary masses along with select orbital parameters. The fit parameters are also summarized in Table \ref{orbit_params}. The primary mass of 1.397$^{+0.050}_{-0.052} M_{\odot}$ is nearly the same as the adopted prior and the secondary mass best fit value is 31$^{+35}_{-11} M_{\rm Jup}$. The companion has a best-fit semimajor axis of 18.6$^{+10}_{-4.1}$ au with an eccentricity of 0.42$^{+0.39}_{-0.29}$ and inclination of 69.4$^{+5.6}_{-15}$ degrees. 

From the corner plot and inset in Figure \ref{fig:cornerplot} it is clear that the low-mass solutions favor less eccentric orbits at shorter semimajor axes. We also note the bimodal behavior of the distribution of eccentricities with peaks at $e\sim0.13$ and $\sim0.81$. Continued monitoring in follow-up observations will serve to further constrain the best-fit values for the orbit of this companion as greater fractions of its orbit are observed. 

%\subsubsection{Choice of Prior}
\subsubsection{Results Using a Gaussian Prior for Companion Mass}

\begin{figure*}
    \centering
    \includegraphics[width=\textwidth]{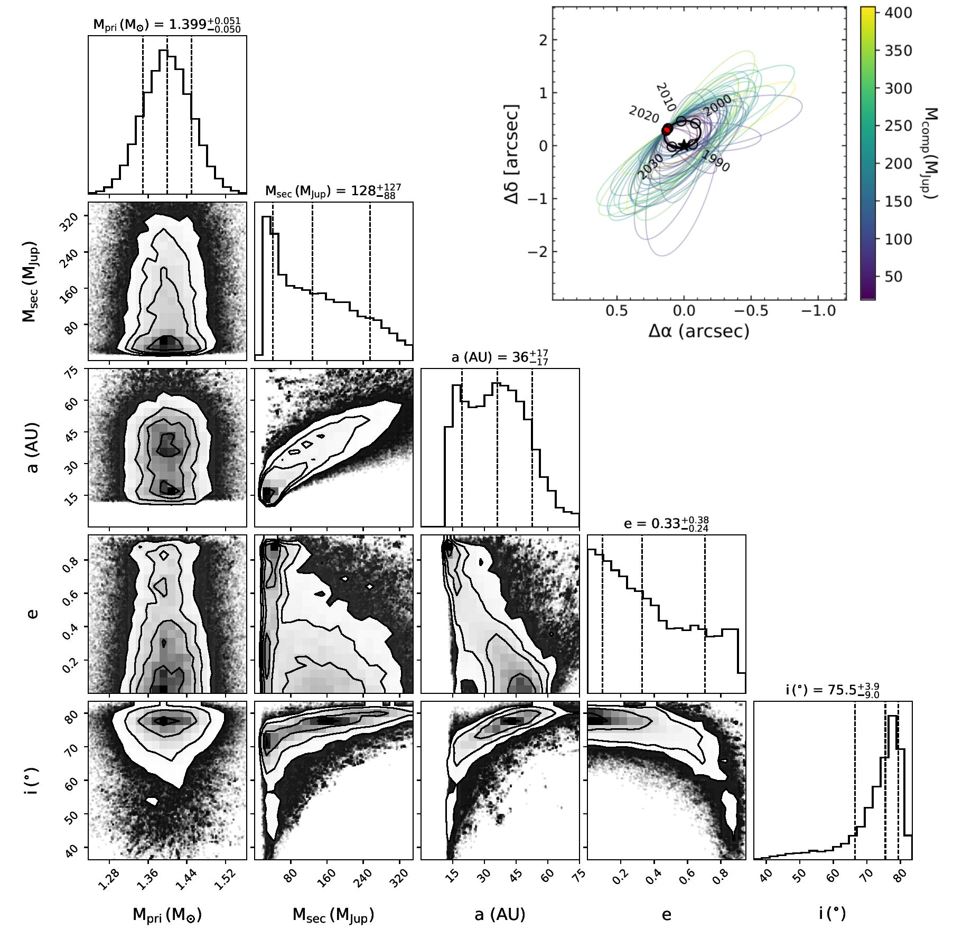}
    \caption{Corner plot showing the posterior distributions of selected orbital parameters using a Gaussian prior of 0.2$\pm0.1 M_{\odot}$ for the mass of the secondary companion. The orbit fits used \textit{Hipparcos} and \textit{Gaia} (HGCA) absolute astrometry and relative astrometry from SCExAO/CHARIS and MEC data. The inset in the figure shows the best fit orbit (black) with 50 random orbits drawn from the MCMC fits color coded by the mass of HIP 5319 B. The red-colored points in the orbit represent relative astrometry points from the 3 epochs where data were taken, and the unfilled circles show the predicted location of the companion at different past and future epochs. The companion is orbiting counterclockwise.}
    \label{fig:cornerplot_gaussian}
\end{figure*}

\begin{deluxetable}{llllll}
    \tablewidth{0pt}
    \tablecaption{HIP 5319 B Orbit Fitting Results For Different Priors on Secondary Mass}
    \tablehead{\colhead{Parameter} & \colhead{log-flat} & \colhead{Gaussian} \\ 
    \colhead{} & \colhead{(1/M)} & \colhead{(0.2$\pm0.1M_{\odot}$)}}
    \startdata
 %   \textbf{New Data}\\
    $M_{pri}$ ($M_{\odot})$ & $1.397^{+0.050}_{-0.052}$ & $1.399^{+0.051}_{-0.050}$ & \\
    $M_{sec}$ ($M_{\rm Jup})$ & $31^{+35}_{-11}$ & $128^{+127}_{-88}$ &  \\
    Semimajor axis \textit{a} (au) & $18.6^{+10}_{-4.1}$ & $36^{+17}_{-17}$ \\
    Eccentricity \textit{e} & $0.42^{+0.39}_{-0.29}$ & $0.33^{+0.38}_{-0.24}$ \\
    Inclination \textit{i} ($^{\circ}$) & $69.4^{+5.6}_{-15}$ & $75.5^{+3.9}_{-9.0}$ \\
    \enddata
    \tablecomments{Posterior distributions for 2 different priors on the secondary mass. The priors on all other parameters being fit remain unchanged between the simulations and can be found for reference in Table \ref{orbit_params}.}
    \label{prior_comparison_params}
    \vspace{-0.5cm}
\end{deluxetable}

We have focused on the \texttt{orvara} fits using a log-flat prior for the secondary mass. 
%but note this may cause concerns of artificially inflating the likelihood of low-mass solutions while attenuating that of high-mass solutions due to the limited constraint of the system dynamics on the mass, causing the posterior to closely match the prior. 
However, the mass function near the hydrogen-burning limit exhibits a turnover, where lower-mass objects are less common \citep{Chabrier2003}. 
%To determine how the choice of prior may affect the posterior distribution for companion mass, we reran \texttt{orvara} using anand to ensure that the choice of prior was not heavily influencing the interpretation of the results we chose a second prior, this one based on inferences made about the companion.}
To investigate how  the choice of prior may affect the posterior distribution for companion mass, we reran \texttt{orvara} using a Gaussian prior of $M_{sec}=0.2\pm0.1 M_{\odot}$ ($210\pm105 M_{\rm Jup}$), comparable to the implied masses for M3-M7 stars (section \ref{sub:spectrum}).  It is also similar to the turnover in the binary mass function from \citet{Chabrier2003}.   Assuming this companion is on the main sequence, the upper limit of its mass would be $M_{sec}\sim0.3-0.4 M_{\odot}$. This prior therefore encompasses these potential values by creating a Gaussian where the expected values of the secondary mass will fall between +2$\sigma$ and -2$\sigma$.

%The resulting best-fit posterior values can be found in Table \ref{prior_comparison_params} \textcolor{red}{and the corresponding corner plot showing these posterior distributions is shown in Figure \ref{fig:cornerplot_gaussian}}. \textcolor{red}{We note here that although the fits which use a Gaussian prior \textit{allow} for higher-mass solutions, the posterior distributions still heavily favor orbital solutions with low-mass companions. The fit parameters are summarized in the final column of table \ref{prior_comparison_params}. The primary mass of $1.399^{+0.051}_{-0.050}$ is again nearly identical to that of the adopter prior. The best-fit semimajor axis is $36^{+17}_{-17}$ au with an eccentricity of $0.33^{+0.38}_{-0.24}$ and inclination of $75^{+3.9}_{-9.0}$ degrees. We note here the bimodal behavior in the values for the semimajor axis with peaks at $a\sim18$ and $\sim36$ au, which will be further constrained by follow-up observations of the system.}

Table \ref{prior_comparison_params} lists the resulting best-fit posterior values; Figure \ref{fig:cornerplot_gaussian} displays the corner plot showing the posterior distributions.   The eccentricity and inclination distributions  -- $e$ = $0.33^{+0.38}_{-0.24}$, $i$ = $75^{+3.9}_{-9.0}$ degrees -- agree with earlier analyses.  However, compared to results for a log-normal companion mass prior, the median of the posterior distributions for HIP 5319 B's mass and semimajor axis have shifted to larger values: $128^{+127}_{-88}M_{\rm Jup}$ and $36^{+17}_{-17}$ au.    For companion mass, the posterior distribution peak is $\sim$20--40 $M_{\rm J}$: comparable to values derived assuming a log-normal companion mass prior.   But the posterior distribution includes a tail of far higher mass solutions, out to $\sim$ 350 $M_{\rm J}$, resulting in a far larger median value.    The semimajor axis posterior distribution contains two peaks -- one near 18 au and a second near 35-40 au.

In practical terms, our analyses are unable to conclusively clarify whether HIP 5319 B is a brown dwarf or a low-mass star.   Dynamical modeling assuming a log-normal companion mass prior favors a brown dwarf at 18.6 au, while modeling adopting a gaussian prior admits a much wider range of companion masses, including those on both sides of the hydrogen burning limit.  The implied masses from masses from atmospheric modeling admit a wide range of possible values: 44 $M_{\rm J}$ to 1675 $M_{\rm J}$.  However, the orbit insets to Figures \ref{fig:cornerplot} and \ref{fig:cornerplot_gaussian} suggest that future astrometric monitoring of HIP 5319 B should clarify the companion's nature.
%posterior distribution
%The primary mass of $1.399^{+0.051}_{-0.050}$ is nearly identical to that of the adopted prior and is consistent with our earlier analyses.

\section{Summary and Discussion}
SCExAO/CHARIS spectroscopy and photometry from SCExAO/MEC, SCExAO/VAMPIRES and Keck/NIRC2 have enabled the identification of a candidate substellar companion to the young F5IV star HIP 5319. Comparisons of the SCExAO/CHARIS spectra to the spectra of objects in the Montreal Spectral Library show this companion to be best matched with M3-M7 dwarfs, with earlier-type M and L dwarfs failing to match the CHARIS spectra measured in $J$ and $K$ bands. By combining measurements from \textit{Hipparcos} and \textit{Gaia} with our relative astrometry from CHARIS/MEC/VAMPIRES/NIRC2 we can constrain the dynamical mass and orbit of HIP 5319 B.   

Assuming a log-normal prior, we find a dynamical mass of $31^{+35}_{-11}M_{\rm Jup}$ for the companion, suggesting that HIP 5319 B is a brown dwarf. The posterior distributions from the fits for dynamical mass show a bimodal distribution in possible eccentricity values, where high-eccentricity solutions are favored at more edge-on inclinations and low-eccentricity solutions are favored for more inclined orbits. However, adopting a Gaussian prior for the companion mass yields a higher mass of 128$^{+127}_{-88}M_{\rm J}$ which favors the interpretation of the companion as a low mass star although the distribution's peak still falls in the substellar range. Future RV measurements, relative astrometry from direct imaging instruments and more precise astrometry from \textit{Gaia} data releases will contribute to further constraining this companion's mass and orbital parameters, providing deeper clarity on this companion's identity.

Atmospheric models of the companion show a best fit to an atmosphere with solar metallicity at T=3200 K with a surface gravity log(g)=5.5, though solutions with comparably good fits exist with temperatures that range from 3100 K to 3300 K and slightly lower surface gravities (log(g)=4-4.5). The best-fit models show radii between 3.25-3.62 $R_{J}$ and log(L/L$_{\rm \odot}$)=$-1.94\pm0.04$. The mass inferred from atmospheric modeling is poorly constrained.

%These values are not consistent with identification of the companion as a brown dwarf unless further study of the system find it to be much younger than previously reported or the companion was formed significantly later than the primary.

This work highlights the need to have an updated inventory of system measurements when interpreting companions imaged around accelerating stars.   While much older data suggested that HIP 5319 is a RS CVn (short-period) binary, our RV data rule out stellar companions with an orbital period less than 30 days whose presence would affect our conclusions about HIP 5319 B's mass and orbital properties.   Similarly, our HIP 5319 spectra find no evidence for Ca HK emission that could reveal evidence of HIP 5319's youth.   Other system measurements whose values may impact derived companion masses and orbits include spectral type/luminosity, projected rotation rate, lithium abundances, x-ray activity, etc.
%clarify HIP 5319's age.    
%we find no evidence From data taken with the NRES instrument via LCOGT we also conclude that the primary does not show signs of variation in RV signals or strong emission in Ca II H $\&$ K lines which allows us to refute earlier claims of HIP 5319 A's identity as an RS CVn binary. Based on our new observations HIP 5319 A is a single star with no strong evidence supporting its identity as a binary. This is of principal importance to the interpretation of the mass of the companion. 

% While its identity as a binary is sufficiently debunked, archival data from TESS and IUE do however show that the primary exhibits pulsations with a period P$\lesssim$1 day and strong emission at Lyman $\alpha$ line. We tentatively suggest that it shares characteristics with Gamma Doradus variable stars, although full identification and characterization of its variability is beyond the scope of this work.

Finally, this work demonstrates the importance of priors in dynamical models used to estimate companion masses and orbits from direct imaging and astrometry.   When a small fraction of a companion's orbit has been observed - as is the case with HIP 5319 B - the selection of prior for a given parameter may influence the final shape of the posterior distributions and the reported values of the dynamical mass and orbital parameters. The chosen prior should not cause the fitted values to change significantly. 
%\citep[see also][]{Currie2022c}
 Performing multiple fits for orbital parameters using disparate priors (e.g. Gaussian, log-normal, uniform, geometric, depending on the parameter of interest) can confirm that the extracted masses and orbital parameters are robust. If the results from multiple fits are in good agreement with one another - the values within the 95$\%$ or 68$\%$ confidence interval overlap with one another, for example - one may say conclusively that the derived dynamical mass is robust. Otherwise, the data are not sufficiently constraining: more of the orbit must then be observed before one can make a definitive claim regarding the fitted orbital parameters and masses of the system.

This direct imaging detection was - in part - made due to the identification of the system as having statistically significant astrometric acceleration in the HGCA. Previous works which include -- but are not limited to -- \citet{Brandt2019}, \citet{Kervella_2019}, \citet{Currie2020}, \citet{Bonavita2020}, \citet{Bowler2021}, \citet{Chilcote2021}, \citet{Li2021}, \citet{Steiger2021}, \citet{Kuzuhara2022}, \citet{Miskovetz2022}, and \citet{Salama2022} have also used the HGCA to select targets that have been found to host previously unidentified companions. This discovery further demonstrates the efficacy of using astrometry to select direct imaging targets instead of conducting blind searches.   As more HGCA targets are observed, future \textit{Gaia} data releases yield more precise astrometry, and direct imaging capabilities improve, this survey approach will only become more powerful in discovering substellar companions, including numerous planets \citep{Currie2021}.
%Ultimately that will continue make this a useful technique for discovering exoplanets via direct imaging. 

\section{Acknowledgments}
 
The authors wish to recognize and acknowledge the very significant cultural role and reverence that the summit of Maunakea has always had within the indigenous Hawaiian community.  We are most fortunate to have the opportunity to conduct observations from this mountain.   

We also wish to thank Eric Mamajek, Jonathan Gagne, and Kris He\l{}miniak for their helpful comments about the HIP 5319 system properties.    We thank the Las Cumbres Observatory for their approval of director's discretionary time (DDT) in order to perform spectroscopic follow-up of the primary under project code DDT2021B-007.   

The development of SCExAO was supported by the Japan Society for the Promotion of Science (Grant-in-Aid for Research \#23340051, \#26220704, \#23103002, \#19H00703 $\&$ \#19H00695), the Astrobiology Center of the National Institutes of Natural Sciences, Japan, the National Astronomical Observatory of Japan, and the Mt Cuba Foundation.

TC was supported by a NASA Senior Postdoctoral Fellowship and NASA/Keck grant LK-2663-948181. TB gratefully acknowledges support from the Heising-Simons foundation and from NASA under grant \#80NSSC18K0439.  MT is supported by JSPS KAKENHI Grant \# 18H05442. NSk and VD acknowledge support from NASA (Grant \#80NSSC19K0336). SS is supported by a grant from the Heising-Simons Foundation. NZ was supported throughout this work by a NASA Space Technology Research Fellowship. KKD is supported by an NSF Astronomy and Astrophysics Postdoctoral Fellowship under award AST-1801983.

%% To help institutions obtain information on the effectiveness of their 
%% telescopes the AAS Journals has created a group of keywords for telescope 
%% facilities.
%
%% Following the acknowledgments section, use the following syntax and the
%% \facility{} or \facilities{} macros to list the keywords of facilities used 
%% in the research for the paper.  Each keyword is check against the master 
%% list during copy editing.  Individual instruments can be provided in 
%% parentheses, after the keyword, but they are not verified.

\vspace{5mm}
\facilities{Subaru/SCExAO, Keck/NIRC2, Las Cumbres Observatory Global Telescopes (LCOGT)}

%% Similar to \facility{}, there is the optional \software command to allow 
%% authors a place to specify which programs were used during the creation of 
%% the manuscript. Authors should list each code and include either a
%% citation or url to the code inside ()s when available.

\software{MKID Data Reduction Pipeline (\citet{Steiger2022}, \citet{Walter2020}), \texttt{orvara} \citep{orvara2021}, CHARIS Data Reduction Pipeline \citep{Brandt2017}, CHARIS Data Processing Pipeline \citep{Currie2020b}, \texttt{matplotlib}, \texttt{astropy}, BANZAI-NRES}

%% Appendix material should be preceded with a single \appendix command.
%% There should be a \section command for each appendix. Mark appendix
%% subsections with the same markup you use in the main body of the paper.

%% Each Appendix (indicated with \section) will be lettered A, B, C, etc.
%% The equation counter will reset when it encounters the \appendix
%% command and will number appendix equations (A1), (A2), etc. The
%% Figure and Table counter will not reset.

% \appendix
% \section{CHARIS Spectra}
% We include the CHARIS spectra from the July 2020 and September 2021 epochs in Table \ref{CHARIS_spectra}.

\bibliography{main_arxiv_post}{}
\bibliographystyle{aasjournal}

%% This command is needed to show the entire author+affiliation list when
%% the collaboration and author truncation commands are used.  It has to
%% go at the end of the manuscript.
%\allauthors

%% Include this line if you are using the \added, \replaced, \deleted
%% commands to see a summary list of all changes at the end of the article.
%\listofchanges

\end{document}